\documentclass[
reprint,           
onecolumn,
superscriptaddress,
amsmath,           
amssymb,           
aps,               
prd,               
notitlepage,       
floatfix,          
nofootinbib,
]{revtex4-1}

\usepackage{amsmath}
\allowdisplaybreaks[4]
\usepackage{cancel}
\usepackage{extarrows}
\usepackage{tensor}     
\usepackage{float}
\usepackage[caption = false]{subfig} 
\usepackage[final]{graphicx}   
\usepackage[
colorlinks=true,        
citecolor=blue,         
linkcolor=blue,         
urlcolor=blue           
]{hyperref}             
\usepackage{bm}         
\usepackage{xcolor}     
\usepackage{lipsum}
\usepackage{color}      
\usepackage[utf8]{inputenc} 
\usepackage[section]{placeins} 
\usepackage{appendix}
\newcommand{\nc}{\newcommand*} 

\nc{\al}{\alpha}
\nc{\bt}{\beta}
\nc{\gm}{\gamma}
\nc{\Gm}{\Gamma}
\nc{\s}{\sigma}
\nc{\Si}{\Sigma}
\nc{\dt}{\delta}
\nc{\Dt}{\Delta}
\nc{\z}{\zeta}
\nc{\Ld}{\Lambda}
\nc{\p}{\partial}
\nc{\om}{\omega}
\nc{\Om}{\Omega}
\nc{\rd}{\mathrm{d}}
\nc{\ma}{\mathcal{A}}
\nc{\mB}{\mathcal{B}}
\nc{\me}{\mathcal{E}}
\nc{\mw}{\mathcal{W}}
\nc{\mx}{\mathcal{X}}
\nc{\mv}{\mathcal{V}}
\nc{\mg}{\mathcal{G}}
\nc{\mh}{\mathcal{H}}
\nc{\mq}{\mathcal{Q}}
\nc{\mm}{\mathcal{M}}
\nc{\Od}[1]{\mathcal{O}(#1)} 
\nc{\kp}{\kappa}
\nc{\dg}{\dagger}
\nc{\na}{\nabla}
\nc{\zt}{\zeta}
\nc{\ko}{\koppa}
\nc{\io}{\iota}
\nc{\ep}{\epsilon}

\def\({\left(}
\def\){\right)}
\def\[{\left[}
\def\]{\right]}	
\def\e{\begin{equation}}
\def\q{\end{equation}}
\def\m{\begin{eqnarray}}
\def\n{\end{eqnarray}}
\def\ali{\begin{align}}
\def\gn{\end{align}}
\nc{\Eq}[1]{Eq.~\eqref{#1}}     
\nc{\Fig}[1]{Fig.~\ref{#1}}     
\nc{\Table}[1]{Table~\ref{#1}}  
\nc{\Sec}[1]{Sec.~\ref{#1}}     
\nc{\Msun}{M_\odot}             
\nc{\fpbh}{f_{\mathrm{pbh}}}    
\nc{\fpbhn}{f_{\mathrm{pbh0}}}    
\nc{\mR}{\mathcal{R}} 
\nc{\seq}{\sigma_{\mathrm{eq}}}
\nc{\ogw}{\Omega_{\mathrm{GW}}}
\nc{\gpcyr}{\mathrm{Gpc}^{-3}\,\mathrm{yr}^{-1}}
\nc{\lvc}{LIGO/Virgo} 
\nc{\SNR}{\mathrm{SNR}} 
\nc{\mmin}{{m_{\mathrm{min}}}}
\nc{\mmax}{{m_{\mathrm{max}}}}
\nc{\Mmin}{{M_{\mathrm{min}}}}
\nc{\fmin}{{f_{\mathrm{min}}}}
\nc{\VT}{\mathrm{VT}}
\nc{\rhoGW}{\rho_{\mathrm{GW}}}
\nc{\vth}{\vec{\theta}}
\nc{\vd}{\vec{d}}
\nc{\vla}{\vec{\lambda}}
\nc{\Nobs}{N_{\mathrm{obs}}}
\nc{\av}[1]{\langle #1 \rangle} 
\nc{\km}{\mathrm{km}}
\nc{\Mpc}{\mathrm{Mpc}}
\nc{\Tobs}{T_{\mathrm{obs}}}
\nc{\Ntemp}{N_{\mathrm{temp}}}
\nc{\addref}{[\textcolor{red}{add ref}] } 
\nc{\eg}{\textit{e.g.~}}
\nc{\app}{\approx}
\nc{\la}{\label}
\nc{\hf}{\frac{1}{2}}
\nc{\discuss}{\textcolor{red}{Add discussion here!}}
\nc{\red}[1]{\textcolor{red}{#1}}
\nc{\po}{|\mathbf{x}-\mathbf{x'}|}
\nc{\ppo}{|\mathbf{x}-\mathbf{x''}|}
\nc{\spo}{|\mathbf{x}-\mathbf{x'}|^3}
\nc{\sspo}{|\mathbf{x}-\mathbf{x'}|^5}
\nc{\pspo}{|\mathbf{x'}-\mathbf{x''}|}
\nc{\nspo}{|\mathbf{x'}-\mathbf{x''}|^3}
\nc{\bv}{\mathbf{v}}
\nc{\bx}{\mathbf{x}}
\nc{\bxp}{\mathbf{x'}}
\nc{\pa}{\partial}
\nc{\rs}{\rho^{*}}
\nc{\eq}{\equiv}
\nc{\nor}{\frac{1}{4\pi}}
\nc{\rp}{\mathrm{P}}
\nc{\rt}{\mathrm{T}}
\nc{\rx}{\mathrm{X}}
\nc{\ry}{\mathrm{Y}}
\nc{\rS}{\mathrm{S}}
\nc{\rsd}{\mathrm{SD}}
\nc{\rw}{\mathrm{W}}
\nc{\rz}{\mathrm{Z}}
\nc{\re}{\mathrm{E}}
\nc{\mH}{\mathcal{H}}
\nc{\cc}[1]{\frac{1}{c^{#1}}}
\nc{\cs}{c_s^2}
\nc{\tl}[1]{\tilde{#1}}
\nc{\Sij}[1]{S_{ij}^{(#1)}}
\nc{\vi}[1]{v_i^{(#1)}}
\nc{\no}{\nonumber}
\def\<{\left\langle}
\def\>{\right\rangle}

\nc{\bk}{\bm{k}}
\nc{\bq}{\bm{q}}
\nc{\bp}{\bm{p}}
\nc{\bl}{\bm{l}}
\nc{\be}{\mathbf{e}}
\nc{\mS}{\mathcal{S}}
\nc{\te}{\tilde{\eta}}
\nc{\tp}{\tilde{p}}
\nc{\tk}{\tilde{k}}
\nc{\tx}{\tilde{x}}
\nc{\tF}{\tilde{F}}
\nc{\tA}{\tilde{A}}
\nc{\mkpq}{|\bk-\bp-\bq|}
\nc{\mpq}{|\bp-\bq|}
\nc{\mkp}{|\bk-\bp|}
\nc{\mSi}[1]{\mS^{(#1)}({\bk, \eta})}
\nc{\vk}{\vec{k}}
\nc{\kstar}{k_*}
\nc{\fstar}{f_*}
\nc{\xstar}{x_*}
\nc{\mpbh}{m_{\rm{pbh}}}
\nc{\bn}[1]{\bm{n}_{\text{#1}}}
\nc{\bC}[1]{\bm{C}_{\text{#1}}}
\nc{\NTOA}{N_{\text{TOA}}}
\nc{\Nmode}{{N_{\text{mode}}}}
\nc{\ARN}{A_{\rm{RN}}}
\nc{\gRN}{\gamma_{\rm{RN}}}
\nc{\bS}{\mathbf{\Sigma}}
\nc{\br}{\mathbf{r}}
\nc{\bN}{\mathbf{N}}
\nc{\arXiv}[2]{\href{http://arxiv.org/pdf/#1}{{\tt [#2/#1]}}}
\nc{\arXivold}[1]{\href{http://arxiv.org/pdf/#1}{{\tt [#1]}}}

\renewcommand{\vec}[1]{\boldsymbol{#1}} 

\begin{document}
	
\title{Parameterized second post-Newtonian framework with conservation Laws}
	
\author{Yu-Mei Wu}
\email{wuyumei@itp.ac.cn}
\affiliation{CAS Key Laboratory of Theoretical Physics, 
	Institute of Theoretical Physics, Chinese Academy of Sciences,
	Beijing 100190, China}
\affiliation{School of Physical Sciences, 
	University of Chinese Academy of Sciences, 
	No. 19A Yuquan Road, Beijing 100049, China}
\author{Qing-Guo Huang}
\email{Corresponding author: huangqg@itp.ac.cn}
\affiliation{CAS Key Laboratory of Theoretical Physics, 
Institute of Theoretical Physics, Chinese Academy of Sciences,
Beijing 100190, China}
\affiliation{School of Physical Sciences, 
University of Chinese Academy of Sciences, 
No. 19A Yuquan Road, Beijing 100049, China}
\affiliation{School of Fundamental Physics and Mathematical Sciences
Hangzhou Institute for Advanced Study, UCAS, Hangzhou 310024, China}
\affiliation{Center for Gravitation and Cosmology, 
College of Physical Science and Technology, 
Yangzhou University, Yangzhou 225009, China}
	
\date{\today}

\begin{abstract}
	
We parameterize the second post-Newtonian (2PN) metric for a gravitating system of fluids in the generalized harmonic gauge, and find that there are only three independent 2PN parameters (i.e. $\omega$, $\delta$ and $\delta_2$) for satisfying some conservation laws including the conservations of energy, momentum, angular-momentum and the uniform motion of the center-of-mass. 


\end{abstract}
	
	
\maketitle


\section{Introduction}
Was Einstein right? Put general relativity (GR) to the test! \cite{Will:1995} The experimental tests prevail in every gravitational observation predicted by GR in different scales. For instance, the solar system provides a terrific weak-field regime \cite{Will:2014kxa,Will:2018bme} where the tests like light deflection and perihelion
advance of Mercury are conducted, and the binary-pulsar as a strong-gravity astrophysical system \cite{Stairs:2003eg} gives access to the comparison between the decrease in its orbital period and gravitational-wave energy loss. In recent years, the thrilling detection of gravitational waves originated from binary black holes \cite{TheLIGOScientific:2016src} and the image of black holes \cite{Akiyama:2019cqa, Akiyama:2019bqs}, through possible direct reflection of the strong field of black holes, mark the advent of a new era for gravitational tests. 

GR has passed all the gravitational tests so far with flying colors, but due to the non-renormalizability of the theory itself and interpretation puzzle for the cosmological constant problem, the efforts to seek its alternatives have never ceased. However, from an observational point of view, in order to incorporate the possible deviation from GR systematically rather than focus on a certain modified gravitational theory, parameterizing gravitational theory from different aspects according to the application range of experiments is an effective method. For instance, the parameterized post-Newtonian (PPN) framework, parameterized post-Keplerian (PPK) framework \cite{PPK1} and parameterized post-Einsteinian (PPE) framework \cite{Yunes:2009ke} have been constructed. Among them, the PPN formalism is a powerful tool governing the realm of weak-field tests.
The idea of PPN framework was originated from Eddington. He parameterized the first post-Newtonian (1PN) limit\footnote{In order to avoid ambiguity, we use ``1PN (order)" to refer to ``the first post-Newtonian (order)", and ``PN (theory)" to refer to ``post-Newtonian (theory) of all orders". However, the traditional notation ``PPN (metric)" specially refers to parameterized post-Newtonian (metric) at the first order. } of a Schwarzschild metric with two arbitrary parameters (namely, $\gm$ and $\bt$) in 1922 \cite{Eddington}, where $\gm$ and $\bt$ can be regarded to separately measure the spatial curvature and nonlinearity produced by gravity. Will and Nordtvedt extended the approach during 1968--1972 by introducing another eight arbitrary parameters in addition to $\gm$ and $\bt$ in front of the independent 1PN potentials to develop the modern version of PPN framework \cite{Nordtvedt:1968qs,Thorne:1970wv,Will:1971zzb,Will:1971wt,Will:1972zz}, where the new parameters are classified according to whether there exist preferred location/frame effects and conserved total momentum. The PPN framework encompasses a large amount of alternative metric theories \cite{Will:2014kxa,Chen:2014qsa,Hohmann:2015kra,Sadjadi:2016kwj} and provides a broad range of testable phenomena. Up to now, the experimental tests performed in the solar system put relatively strong constraints on the PPN parameters and therefore on the modified gravitational theories \cite{Will:2014kxa}. 

After the success of the PPN framework, the attempts to generalize the parameterization to the second post-Newtonian (2PN) order were made from different directions. For example, the Schwarzschild metric is  directly expanded to 2PN order and then parameterize the spatial-spatial component $g_{ij}$ to account for a future more accurate measurement of the light deflection \cite{Epstein:1980dw, Deng:2015aga}. A more ambitious effort from Nordtvedt and Benacquista tried to include any possible metric theory at this order and has obtained the independent parameters on the requirement of Lorentz invariance of Lagrangian \cite{B&N1,B2,N3}. But Demour pointed out that these parameters are not indeed constants and introduced a field-theory-based framework to find out that only two independent 2PN parameters are needed in the ``tensor-multiscalar" theories \cite{Damour:1995kt}. Thereafter, several specific modified theories have also been calculated to their 2PN approximation  \cite{Xie:2007gq, Xie:2008ry}, but a parameterized second post-Newtonian (P2PN) theory parallel to PPN theory has not been actually constructed.

On one hand, the traditional weak field tests is expected to obtain unprecedented precision with present and future astrometric missions, such as Gaia \cite{Perryman:2001sp} and Laser Astrometric Test of Relativity (LATOR) \cite{Turyshev:2003wt}, to ${\cal O}(10^{-9}\sim 10^{-7})$ of the size of relativistic effects, which is beyond the level of 1PN effects $(G\Msun/(R_\odot c^2)\sim {\cal O}(10^{-6}))$ and requires taking the 2PN effects into account.  
On the other hand, the information of a 2PN parameter also resides in the strong field in a parameterized Kerr metric \cite{Johannsen:2011dh}, and can be extracted with present and future gravitational wave test \cite{Carson:2020iik} and the horizon-scale images of black hole \cite{Psaltis:2020lvx}. The reason is the n-th post-Newtonian approximation acts as the asymptotic form of a black hole metric. Actually, the parameterized Kerr metrics in \cite{Johannsen:2011dh,Konoplya:2016jvv} match with the PPN metric in the weak-field region \cite{Johannsen:2011dh,Konoplya:2016jvv}. If we require a higher order accuracy, in order to keep the compatibility of parameterization in the weak-field and strong-field regions, a reasonable P2PN formalism is also necessary. 

In this paper, we will mainly follow Will's approach to construct a reasonable P2PN framework. However, it is rather difficult to seek out all the possible 2PN potentials, especially the ones absent in the 2PN limit of GR. Through a brief analysis on the PPN framework, we find a compromise to simplify the P2PN structure and concentrate on searching the independent P2PN parameters under fairly stringent constraints on the metric theory; that is, the theory should contain no preferred location/frame effects and satisfy the conservation laws. Our paper is organized as follows. We give a brief review on the PPN framework in Section \ref{sec2}. In Section \ref{sec3}, we construct a primitive P2PN metric and establish the gravitational stress-energy pseudotensor with the help of some transformation tricks and obtain the parameter constraints. The gauge that the metric with the constraints satisfies is also discussed. Finally, we give a summary and a brief discussion in Section \ref{sec4}. Some details are added in Appendix \ref{app1} and \ref{app2}.

We adopt the following conventions and notations in the paper. The signature of metric takes $(-,+,+,+)$. Greek indices take the values from 0 to 3, while Latin indices
take the values from 1 to 3. Einstein's summation rule over repeated indices is used, even when we only have spatial indices and are not careful about raising and lowering the indices. Parentheses around the indices indicate the indices being symmetrized and square brackets indicate anti-symmetrization; for example, $A_{(i}B_{j)}=\hf(A_iB_j+A_jB_i)$ and $A_{[i}B_{j]}=\hf(A_iB_j-A_jB_i)$. Bold letters $\mathbf{v}=v^{i}$ denote spatial vectors. 
Since the post-Newtonian expressions of GR are essentially the retarded solutions to wave equations expanded in the near-zone field and involve Poisson-like potentials and their generalizations, here we take the definitions of the Poisson potential, superpotential and superduperpotential for a soure $f$ following Will's notation in \cite{Pati:2000vt}, 
\m
\rp(f)&\eq& \nor \int \frac{f(t,\bx')}{\po} d^3x',\qquad \qquad \qquad \quad \na^2 \rp(f)=-f,\notag\\
\rS(f)&\eq& \nor \int f(t,\bx')\po d^3x',\qquad \qquad\, \na^2 \rS(f)=2\rp(f),\notag\\
\rsd(f)&\eq& \nor \int f(t,\bx')\spo d^3x',\!\qquad \qquad \na^2 \rsd(f)=12\rS(f).
\label{Sdp}
\n
We also set $G=1$ throughout the paper, where $G$ the gravitational constant.

\section{ A brief review of the parameterized Post-Newtonian framework }\label{sec2}

Before constructing the P2PN formalism, we give a brief review on the PPN framework and delineate the method for parameterization of metric.

For a self-gravitating fluid system with the energy-momentum tensor expressed by the proper mass density $\rho$, the internal energy per unit mass $\Pi$, the pressure $P$ and the four-velocity field $u^{\al}$ as follows 
\m
T^{\al\bt}=\(\rho(1+\Pi/c^2)+P/c^2\)u^\al u^\bt +P g^{\al\bt},
\label{Tab1}
\n
the PPN metric in the generalized standard gauge is 
\m
g_{00}&=&-1+\frac{2}{c^2}U+\frac{1}{c^4}(-2 \bt U^2+3\bt_1\Phi_1-2\bt_2\Phi_2+2\bt_3\Phi_3+6\bt_4\Phi_4) +\frac{1}{c^4}(\bt_6 \Phi_6 +\xi \Phi_W+\Phi^{\rm{PF}})+\Od{c^{-6}},\notag \\
g_{0j}&=&-\frac{4}{c^3}\Dt V_{j}-\frac{1}{2c^3}\Dt_1 \pa_{tj}X+\frac{1}{c^3}\Phi^{\rm{PF}}_j+\Od{c^{-5}},\notag \\
g_{jk}&=&(1+\frac{2}{c^2}\gm U)\dt_{jk}+\Od{c^{-4}}.
\label{S1P}
\n 
Here $U$ is the Newtonian gravitational potential given by
\m
U=U(\bx,t)\equiv \int \frac{\rho^{*}(t,\mathbf{x}')}{\po}d^3 x'=\int \frac{\rho^{*}{'}}{\po}d^3 x'=\rp(4\pi \rs),
\n
where $\rho^{*}$ is the rescaled mass density related to the proper mass density $\rho$ by $\rs=\sqrt{-g} \rho u^0/c$. The rescaled mass density is also called the conserved mass density because it satisfies the continuity equation, \cite{Will:2014sgx},
\m
\pa_{t} \rs+\pa_{j}(\rs v^{j})=0, 
\label{CE}
\n 
where $v^j$ is a three-velocity field in $u^\al$, namely $u^{\al}=(u^0/c)(c,\bv)$. Some other potentials in \Eq{S1P} are
\m
\Phi_1&\equiv&\rp(4\pi \rs v^2), \quad \qquad 
\Phi_2\equiv\rp(4\pi \rs U),\quad \qquad 
\Phi_3\equiv\rp(4\pi \rs \Pi),\quad \qquad 
\Phi_4\equiv\rp(4\pi P),\notag\\
V^{j}&\equiv& \rp(4\pi \rs v^j),\quad \qquad 
X\eq\int \rho^{*}{'}\po d^3 x'=\rS(4 \pi \rs),
\notag\\
\Phi_6&\equiv&\int \frac{\rho^{*}{'}v_j'v_k'(x-x')^j(x-x')^k}{\spo}d^3 x',\qquad 
\Phi_W\equiv\int \frac{\rho^{*}{'}\rho^{*}{''}(x-x')_j}{\po^3}\[\frac{(x'-x'')^j}{\ppo}-\frac{(x-x'')^j}{\pspo}\]d^3 x'd^3 x'' ;
\label{PNP}
\n
and the rest are preferred-frame potentials,
\m
\Phi^{\rm{PF}}=\al_1 w^2 U+\al_2w^jw^k \pa_{jk}X+\al_3w^j V_j,\qquad\qquad
\Phi^{\rm{PF}}_j=\al_4w_j U+\al_5w^k\pa_{jk}X,
\n
where $w^j$ is the velocity of the PPN coordinate system relative to the universal preferred frame. We make some clarifications in regard to the metric in \Eq{S1P}:\\
(1)  When $\bt=\bt_1=\bt_2=\bt_3=\bt_4=\Dt=\Dt_1=\gm=1$ and all other parameters vanish, the metric reduces to the 1PN metric in GR.\\
(2) The additional potentials absent in GR, i.e. $\Phi_6$, $\Phi_W$, $\Phi^{\rm{PF}}$ and $\Phi^{\rm{PF}}_j$, are related to preferred-location/frame effects. $\Phi_6$ and $\Phi_W$ appear in theories with preferred-location effects caused by, for example, a galaxy-induced anisotropy \cite{Will:2014kxa}, while $\Phi^{\rm{PF}}$ and $\Phi^{\rm{PF}}_j$ are involved in some well-motivated vector-tensor \cite{Foster:2005dk} and tensor-vector-scalar \cite{Sagi:2009kd} theories that include a dynamical timelike vector field, such as Einstein-Æther theory.\\
(3) A metric theory based on an invariant action principle automatically admits a conservation law for total momentum and is called ``semi-conservative". Furthermore, if the theory also admits a conserved angular momentum and a center-of-mass with a uniform motion (which means the existence of a preferred frame is impossible due to the break of local Lorentz invariance), it is called ``fully-conservative".  For a fully-conservative metric theory without preferred-location effects,  only two independent parameters  $\bt, \gm$ retain, and the other non-vanishing parameters including $\bt_1, \bt_2, \bt_3, \bt_4, \Dt, \Dt_1$ are all determined by $\bt, \gm$. Scalar-tensor theories \cite{Damour:1992we}, such as Brans-Dicke theory, fall into this category.\\
So we conclude that in a fully-conservative PPN theory without preferred effects ($\bt_6=\xi=\al_1=\al_2=\al_3=\al_4=\al_5=0$), the potentials are just the same as those in GR but with the coefficients in front of them replacing by the 1PN parameters; furthermore, we can extract the independent ones $\gm$ and $\bt$ from all these 1PN parameters.

\section{The Parameterized Second Post-Newtonian Formalism}\label{sec3}

In this section, we extend the former parameterization to 2PN order, or equivalently we  need to expand the metric to the following order, \cite{Pati:2000vt},
\m
g_{00} \text{\quad to\quad} \Od{c^{-6}},\qquad g_{0j} \text{\quad to\quad}\Od{c^{-5}},\qquad g_{jk} \text{\quad to\quad}\Od{c^{-4}}.
\label{2nd}
\n
In principle, the general P2PN framework can be constructed parallely to that of PPN in the original paper \cite{Will:1971zzb} through listing all the possible 2PN potentials and parametrizing them with arbitrary parameters, and then choosing a gauge to get rid of the redundant degrees of freedom (non-independent potentials). However, the difficulties also reside in two sides correspondingly: one is to find potentials absent in GR, similar to $\Phi_W$ in the PPN (see \Eq{PNP}), that need elaborate construction; the other is to choose a suitable gauge that is compatible from the 1PN order to the 2PN order since the gauge-fixing process involves both orders. 

Therefore, instead of trying to build the general P2PN framework once and for all, we focus on seeking the independent P2PN parameters in a fully-conservative theory without preferred-location/frame effects first. From the discussion in Sec.~\ref{sec2}, this can be achieved by first parameterizing the 2PN metric in GR in a specific gauge, and then figuring out the constraints on these parameters by considering some conservation laws. Such a method of parametrization will certainly break the gauge that stands in GR, but we may find the corresponding gauge for the parameterized metic that meets the parameter constraints hopefully.

\subsection{The P2PN metric}\label{sec3a}

We need to note that the choice of generalized standard gauge of the PPN metric in \Eq{S1P} is only a reflection of historical development based on Chandrasekhar's work \cite{Chandrasekhar:1965gcg} on the classic approach to post-Newtonian theory. By contrast, the harmonic gauge prevails in the modern approach to PN theory, and hence we choose to parameterize the 2PN metric in the harmonic gauge, \cite{Pati:2000vt,Pati:2002ux}\footnote{A systematic method to calculate the post-Newtonian approximation to 3.5 order for unspecified matter fields is given in \cite{Pati:2000vt}, and the matter fields are only consisting of baryons in \cite{Pati:2002ux}. Here, we need to generalize the calculation to perfect fluid to 2PN order in the Appendix \ref{app1}.},
\m
g_{00}&=&-1+\frac{2}{c^2}U+\frac{1}{c^4}\(-2\bt U^2+\Psi_1\)+\frac{1}{c^6}(\frac{4}{3}\dt U^3+ U \Psi_2+8\lambda V^2+\me+\mx+\mg+\mh),\notag\\
g_{0j}&=&-\frac{4}{c^3}\Dt V^j+\frac{1}{c^5}(\chi UV^j+\mv^{j}),\notag\\
g_{jk}&=&\[1+\frac{2}{c^2}\gm U+\frac{1}{c^4}(2\om U^2+\Psi_3)\]\dt_{jk}+\frac{1}{c^4}\mm_{jk},
\label{M2pn}
\n
in which,
\m
&&\Psi_1\eq  3\bt_1\Phi_1-2\bt_2\Phi_2+2\bt_3\Phi_3+6\bt_4\Phi_4 +\bt_0\ddot{X},\notag\\
&&\Psi_2\eq-6 \text{$\delta_1$} \text{$\Phi_1 $}+4 \text{$\delta_2 $} \text{$\Phi_2 $}-4 \text{$\delta_3 $} \text{$\Phi_3 $}-12 \text{$\delta_4 $} \text{$\Phi_4 $}-2\dt_0 \ddot{X},\notag\\
&&\me \eq 3\kp \Si(U^2)-3\kp_1\Si(\Phi_1)-2\kp_2\Si(\Phi_2)-2\kp_3\Si(\Phi_3)+2\kp_4\Si(\Phi_4)-\kp_0\Si(\ddot{X})\notag\\
&&\qquad+\frac{7}{4}\s_1 \Si(v^4)+9\s_2 \Si(v^2 U)-8\s_3\Si(v_j V^j)+3\s_4 \Om(v^2)+4\s_5 \rt(v^2)-2\s_6\Om(U)+12\s_7\rt(U)\notag\\
&&\mx\eq\frac{3}{2}\mu_1\ddot{X_1}-\mu_2\ddot{X_2}+\mu_3\ddot{X_3}+3\mu_4\ddot{X_4}+\frac{1}{12}\mu_0\mathop{\ry}\limits^{(4)}\notag\\
&&\mg\eq-12\xi_1 G_1-8\xi_2 G_2+16\xi_3 G_3+16\xi_4 G_4+\xi_5 G_5,\notag\\
&&\mh\eq-8\eta_1 H_1-8\eta_2H_2,\notag\\
&&\mv^{j}\eq-2\chi_1 V_1^{j}-4\chi_2 V_2^{j}-4\chi_3 V_3^{j}-4\chi_4 V_4^{j}+8\chi_5 \phi_2^{j}-2\chi_0 \ddot{X}^{j}-16\iota_1 K_1^{j}-12\io_2 K_2^{j},\notag\\
&&\Psi_3 \eq -\om_1\Phi_1-2\om_2\Phi_2+2\om_3\Phi_3-2\om_4\Phi_4 +\om_0\ddot{X},\notag\\
&&\mm^{jk} \eq 4\theta_1\Phi_1^{jk}+4\theta_2 P_2^{jk},
\label{P2pn}
\n
where the overdot denotes the derivative with respect to time coordinate and $\mathop{\ry}\limits^{(4)}=\pa_t^{(4)}Y$.
Here most of the potentials in Eqs.~(\ref{M2pn}) and (\ref{P2pn}) can be conveniently classified into several particular classes given in \Eq{Sdp}, namely
\m
\Si(f)&\eq& \int \frac{\rs{'}f'}{\po} d^3x'=\rp(4\pi \rs f),\qquad \qquad \quad \Si^j(f)\eq \int \frac{\rs{'}v'^j f'}{\po} d^3x'=\rp(4\pi \rs v^j f), \notag \\ 
\Si^{ij}(f)&\eq& \int \frac{\rs{'}v'^i v'^j f'}{\po} d^3x'=\rp(4\pi \rs v'^i v^j f), \notag\\
\rx(f)&\eq&  \int \rs{'} f'\po d^3x'=\rS(4\pi\rs f),\qquad \quad \, \rx^j(f)\eq  \int \rs{'}v'^j f'\po d^3x'=\rS(4\pi\rs v'^j f),\notag\\
\ry(f)&\eq&  \int\rs{'}f'\spo d^3x'=\rsd(4\pi \rs f),
\label{Spc}
\n
and then the unspecified potentials can be defined by 
\m
&& V_1^{j}\eq \Si^{j}(v^2),\quad\,\, V_2^{j}\eq \Si^{j}(U),\quad\,\, V_3^{j}\eq \Si^{j}(\Pi),\quad\,\, V_4^{j}\eq
\Si^{j}(P/\rs),\notag\\
&& \phi_2^{j}\eq \Si(V^{j}),\quad\,\, {X}^j \eq \rx^j(1),\quad\,\, K_1^j\eq\rp(U^{,k}V^{k,j}),\quad\,\, K_2^j\eq\rp(U^{,j}\dot{U}),\notag\\
&& \Phi_1^{ij}\eq \Si^{ij}(1),\quad\,\, P_2^{ij}\eq \rp(U^{,i}U^{,j}),\quad  P_2\eq P_2^{ii}=\Phi_2-\hf U^2,\notag\\
&&\Om(v^2)\eq \Si(\Pi v^2),\quad\,\, \rt(v^2)\eq \Si(P/\rs v^2),\quad\,\, \Om(U)\eq \Si(\Pi U),\quad\,\, \rt(U)\eq \Si(P/\rs U),\notag\\
&& X_1\eq\rx(v^2),\quad\,\, X_2\eq\rx(U),\quad\,\, X_3\eq\rx(\Pi),\quad\,\, X_4\eq\rx(P/\rs),\quad\,\,
Y\eq\ry(1),\notag\\
&& G_1\eq \rp(\dot{U}^2),\quad\,\, G_2\eq\rp(U \ddot{U}),\quad\,\, G_3\eq-\rp(\dot{U}^{,k}V^k),\quad\,\, G_4\eq\rp(V^{i,j}V^{j,i}),\quad\,\,G_5\eq-\rp(\dot{V}^k U^{,k}) \notag\\
&& H_1\eq\rp(\Phi_1^{ij}U^{,ij}),\quad\,\, H_2\eq\rp(P_2^{ij}U^{,ij}),
\n
where the commas in the definition denote the partial derivatives with respect to the chosen coordinates.
Note that terms of half-odd-integer PN order, i.e. 1.5 PN order exactly in $g_{00}$, representing the dissipative radiation-reaction, are not taken into account. \Eq{M2pn} recovers the 2PN metric in GR if $\chi=\xi_5=0$ and the other parameters in Eqs.~(\ref{M2pn}) and (\ref{P2pn}) are all equal to 1.

\subsection{The conservation laws and parameter constraints}\label{sec3b}

Even though a large number of parameters are introduced in the general metric \Eq{M2pn}, many of them are not actually independent once the metric is required to satisfy some  conservation laws. 

It is well known, for example in \cite{Landau:1982dva}, that the usual version of energy-momentum  conservation expressed by the vanishing of convariant divergence of $T^{\al\bt}$,
\m
\nabla_{\bt}T^{\al\bt}=0,
\label{VCD}
\n
cannot afford the global conserved quantities since it is a direct consequence of local conservation of energy-momentum ($T^{\al\bt}$) and the contribution from gravitational fields is implicit. On the other hand, another version compatible with \Eq{VCD} expressed by the vanishing of ordinary divergence of a pseudotensor $\Theta^{\al\bt}$, namely 
\m
\pa_{\bt}\Theta^{\al\bt}=0,
\label{VOD}
\n  
can be exploited to determine the conserved quantities. Here an appropriate $\Theta^{\al\bt}$ should follow two rules: (1) it contains contributions from both the matter fields and the gravitational fields; and (2) it reduces to $T^{\al\bt}$ in the flat spacetime. For example, in GR, the exact version of $\Theta^{\al\bt}$ takes $\Theta^{\al\bt}=(-\tilde{g})(T^{\al\bt}+t_{LL}^{\al\bt})$, where $\tilde{g}$ is the determinant of the metric, and $t_{LL}^{\al\bt}$, the famous Landau-Lifshitz pseudotensor, acts as the contribution from the gravitational fields \cite{Landau:1982dva}. Then, the total momentum over a region $\mathcal{V}$ can be formally defined by the three-dimensional integral, \cite{Will:2014sgx},
\m
P^\al:=\frac{1}{c}\int_\mathcal{V} \Theta^{\al 0} d^3 x.
\n
When we take the limit of $\mathcal{V}$ to include all of the three-dimensional space of an asymptotically-flat spacetime, $P^\al$ can be proven to be invariant over time by use of \Eq{VOD}. Similarly, the total angular momentum can be formally defined by 
\m
J^{\al\bt}:=\frac{2}{c}\int_\mathcal{V} x^{[\al}\Theta^{\bt]0} d^3 x 
\n
which is conserved over all of the three-dimensional space only when $\Theta^{\al\bt}$ is also symmetric. More specifically, the limit of $P^\al$ and $J^{\al\bt}$ of an asymptotically-flat spacetime are identified : $P^0$ is the total energy, $P^i$ the total three-momentum, $J^{ij}$ the total angular-momentum three-tensor  and $J^{0j}$ determines the motion of the center-of-mass.

From now on, the metric in \Eq{M2pn} is considered to be fully conservative, or equivalently we need to find the corresponding symmetric $\Theta^{\al\bt}$ satisfying \Eq{VOD}. In the PPN formalism,  such a $\Theta^{\al\bt}$ assumed to be $(1+c^{-2} A U)(T^{\al\bt}+t^{\al\bt})$ has been found by determining the constant $A$ and the symmetric gravitational stress-energy pseudotensor $t^{\al\bt}$ with the use of Eqs. (\ref{VCD}) and (\ref{VOD}) in \cite{Will:1971wt}. When we extend to the P2PN case and follow the rules for an appropriate $\Theta^{\al\bt}$, without loss of generality, we assume
\m
\Theta^{\al\bt}=\(1+\frac{1}{c^2}A U+\frac{1}{c^4} (B U^2+\Psi_4)\)(T^{\al\bt}+t^{\al\bt}),
\label{TET}
\n
with 
\m
\Psi_4=B_1\Phi_1+B_2\Phi_2+B_3\Phi_3+B_4\Phi_4+B_0\ddot{X},
\n
where $A$, $B$ and $B_k$ ($k=0,1,...,4$) are constants. We need to point out that  $t^{\al\bt}$ acting as the contribution from gravitational fields should be constructed from various gravitational potentials (e.g., $U$, $V^j$ and potentials contained in \Eq{P2pn}) and their derivatives, but not contain any of the fluid variables ($\rs$, $v^j$, $\Pi$ and $P$) explicitly.

Substituting \Eq{TET} into \Eq{VOD}, and utilizing \Eq{VCD} which can be rewritten by 
\m
0=\nabla_\bt T^{\al\bt}=\pa_\bt T^{\al\bt}+\Gm^\al_{\mu\bt}T^{\mu\bt}+\Gm^{\bt}_{\mu\bt}T^{\al\mu},
\label{POT}
\n
we convert \Eq{VOD} for $\Theta^{\al\bt}$  to the equation for $t^{\al\bt}$, 
\m
&&\pa_{\bt}\[\(1+\frac{1}{c^2}A U+\frac{1}{c^4} (B U^2+\Psi_4)\)t^{\al\bt}\]\notag\\=&&\(1+\cc{2} A U+\frac{1}{c^4} (B U^2+\Psi_4)\)(\Gm^\al_{\mu\bt}T^{\mu\bt}+\Gm^{\bt}_{\mu\bt}T^{\al\mu})-\(\cc{2}A\pa_\bt U+\cc{4}\pa_\bt(B U^2+\Psi_4)\)T^{\al\bt},
\label{PF}
\n
where the energy-momentum tensor to the required order are given by
\m
c^{-2}T^{00}=&&\rs\left\{1+\frac{1}{c^2}\[\frac{1}{2}v^2+(2-3\gm)U+\Pi\]+\frac{1}{c^4}\[\frac{3}{8}v^4+(2-\frac{1}{2}\gm)U v^2+\frac{1}{2}\Pi v^2+P/\rs v^2-4\Dt v_jV^j\notag\right.\right.\notag\\
&&\left.\left.+(4-2\bt-6\gm+\frac{15}{2}\gm^2-3\om)U^2+(2-3\gm)U\Pi+\Psi_1-\frac{3}{2}\Psi_3-\frac{1}{2}\mm\]\right\}\notag\\
c^{-1}T^{0j}=&&\rs v^j\left\{1+\frac{1}{c^2}\[\frac{1}{2}v^2+(2-3\gm)U+\Pi+P/\rs\]+\frac{1}{c^4}\[\frac{3}{8}v^4+(2-\frac{1}{2}\gm)U v^2+\frac{1}{2}\Pi v^2+P/\rs v^2-4\Dt v_jV^j\notag\right.\right.\notag\\
&&\left.\left.+(4-2\bt-6\gm+\frac{15}{2}\gm^2-3\om)U^2+(2-3\gm)U\Pi+2UP/\rs+\Psi_1-\frac{3}{2}\Psi_3-\frac{1}{2}\mm\]\right\}\notag\\
&&-\frac{4}{c^4}\Dt P V^{j}\notag\\
T^{jk}=&&\rs v^j v^k\left\{1+\frac{1}{c^2}\[\frac{1}{2}v^2+(2-3\gm)U+\Pi+P/\rs\]+\frac{1}{c^4}\[\frac{3}{8}v^4+(2-\frac{1}{2}\gm)U v^2+\frac{1}{2}\Pi v^2+P/\rs v^2-4\Dt v_jV^j\notag\right.\right.\\
&&\left.\left.+(4-2\bt-6\gm+\frac{15}{2}\gm^2-3\om)U^2+(2-3\gm)U\Pi+2UP/\rs+\Psi_1-\frac{3}{2}\Psi_3-\frac{1}{2}\mm\]\right\}\notag\\
&&+P\left\{1-\frac{2}{c^2}\gm U+\frac{1}{c^4}\[(4\gm^2-2\om)U^2-\Psi_3\]\right\}\dt^{jk}-\frac{1}{c^4} P \mm^{jk},
\label{Tab}
\n
with $\mm\equiv\mm^{ll}$,
and the Christoffel symbols are given by
\m
\Gm^0_{00}=&&-\frac{1}{c^3}\pa_{t}U+\frac{1}{c^5}\[(-2+2\bt)U\pa_t U-\frac{1}{2}\pa_t\Psi_1+4\Dt V^l\pa_l U\],\notag\\
\Gm^0_{0j}=&&-\frac{1}{c^2}\pa_{j}U+\frac{1}{c^4}\[(-2+2\bt)U\pa_j U-\frac{1}{2}\pa_j\Psi_1\],\notag\\
\Gm^0_{jk}=&&\frac{4}{c^3}\Dt\pa_{(j}V_{k)}+\frac{1}{c^5}\[\hf\pa_t\mm_{jk}+(-8\gm\Dt-\chi) V_{(j}\pa_{k)}U+(8\Dt-\chi) U\pa_{(j}V_{k)}-\pa_{(j}\mv_{k)}\]\notag\\
&&+\left\{\cc{3} \gm\pa_t U+\cc{5}\[(2\gm+2\om)U\pa_t U+\hf \pa_t \Psi_3+4\gm \Dt V^l\pa_l U\]\right\}\dt_{jk}\notag\\
\Gm^j_{00}=&&-\cc{2}\pa_j U+\cc{4}\[-4\Dt \pa_t V_{j}+(2\bt+2\gm)U\pa_j U-\hf \pa_j \Psi_1\]\notag\\
&&+\cc{6}\[(-4\Dt+\chi) V^j\pa_t U+(8\gm\Dt+\chi) U\pa_t V^{j}+\pa_t \mv^j-8\lambda V^k \pa_j V^k+(-4\bt\gm-4\gm^2-2\dt+2\om)U^2\pa_j U\right.\notag\\
&&\left.+(-\hf \Psi_2+\Psi_3)\pa_j U+\mm^{jk}\pa_kU+U\pa_j(\gm \Psi_1-\hf \Psi_2)-\hf\pa_j(\me+\mx+\mg+
\mh)\]\notag\\
\Gm^j_{0k}=&&\frac{4}{c^3}\Dt\pa_{[j}V_{k]}+\cc{5}\[\hf \pa_t \mm_{jk}-4\Dt V^j\pa_k U+\chi V_{[j}\pa_{k]} U +(-8\gm\Dt-\chi) U\pa_{[j}V_{k]} -\pa_{[j}\mv_{k]}\]\notag\\
&&+\left\{\cc{3}\gm\pa_t U+\cc{5}\[(-2\gm^2+2\om)U\pa_t U+\hf\pa_t \Psi_3\]\right\}\dt_{jk}\notag\\
\Gm^j_{kn}=&&\cc{2}\gm(\dt_{jn}\pa_k U+\dt_{jk}\pa_n U-\dt_{kn}\pa_j U)+\cc{4}\left\{\hf\pa_k \mm_{jn}+\hf\pa_n \mm_{jk}-\hf\pa_j \mm_{kn}\right.\notag\\&&\left.+\dt_{jn}\pa_k \[(-\gm^2+\om)U^2+\hf\Psi_3\]+\dt_{jk}\pa_n \[(-\gm^2+\om)U^2+\hf\Psi_3\]-\dt_{kn}\pa_j \[(-\gm^2+\om)U^2+\hf\Psi_3\]\right\}.
\label{Cabc}
\n 
In principle, the form of a symmetric $t^{\al\bt}$ and the constants $A$, $B$ and $B_k$ ($k=0,1,...,4$) can be obtained by solving \Eq{PF}. In following part of this subsection, we only work out the constraints on the parameters by taking into account the conservation laws, but not the explicit form of $t^{\al\bt}$.

We notice that \Eq{PF} is to turn the quantities involving the fluid variables on the right-hand side into a combination of gradients and time derivatives of gravitational fields on the left-hand side.
In order to solve \Eq{PF}, several transformation tricks would be used. For example, we take advantage of \Eq{Sdp} and also \Eq{Spc} in a reverse direction to convert the various source into field quantities,
\m
-f=\na^2 \rp(f)=\pa_k\pa_k \rp(f),\quad 
-4\pi \rs f= \pa_k\pa_k \Si(f), \quad
-4\pi \rs v^j f=\pa_k\pa_k \Si^j(f),\quad -4\pi \rs v^i v^j f=\pa_k\pa_k \Si^{ij}(f),
\label{STF}
\n
and use the following transformation identity valid for any potential $\Phi$,
\m
-f \pa_\al \Phi&=&-\pa_\al[\pa_k \Phi \pa_k \rp(f)]+\pa_k[\pa_\al \Phi \pa_k \rp(f)]+\pa_k \Phi \pa_k\pa_\al \rp(f),\notag\\
\text{or} \quad
-f \pa_\al \Phi&=&-\pa_\al[\pa_k \Phi \pa_k \rp(f)]+2\pa_k[\pa_{(\al} \Phi \pa_{k)} \rp(f)]-\pa_\al \rp(f) \na^2 \Phi.
\label{TIP}
\n
Some identities between potentials,
\m
&&\pa_t U+\pa_j V^j=0, \label{PI1}\\
&&\pa_t U^2\!+\pa_j(UV^j\!+V_2^j-\phi_2^j\!\!+2K_1^j\!+2K_2^j)=0,\label{PI2}\\
&&\pa_t (7U^2\!+2\Phi_1\!-\!2\Phi_2\!+4\Phi_3\!+2\ddot{X})\!+\pa_j(8UV^j\!+2V_1^j\!+4V_2^j\!+4V_3^j\!+4V_4^j\!-\!8\phi_2^j\!+2\ddot{X}^j\!+16K_1^j\!+12K_2^j)=\Od{c^{-2}},\label{PI3}\\
&&\pa_t V^j+\pa_i \[4\Phi_1^{ij}+4P_2^{ij}-\dt^{ij}(U^2-2\Phi_2-4\Phi_4)\]=\Od{c^{-2}},
\label{PI4}
\n 
are also useful. The identities in Eqs. (\ref{PI1}) and (\ref{PI2}) are actually the results of the continuity equation \Eq{CE}, because the continuity equation \Eq{CE} leads to $\pa_t F=\int \rho^{*}{'}(\pa_t h+v'^{k}\pa_{k'}h)d ^3 x'$ for any function $F(t,x)=\int \rho^{*}{'}h(t,\bx,\mathbf{x}')d^3 x'$ defined by $\rs$. The third identity in \Eq{PI3} and the fourth in \Eq{PI4} are equivalent to the conservation equation of Newtonian total energy (the sum of kinetic, gravitational, and internal energies) \cite{Will:1971wt} and the Euler's equation, respectively,  
\m
&&\rs\pa_t\(\hf v^2+\Pi\)+\rs v^j\pa_j\(\hf v^2+\Pi\)+\pa_j(Pv^j)-\rs v^j \pa_j U=\Od{c^{-2}},\notag\\
&&\rs\frac{d v^j}{dt}=\rs \pa_j U-\pa_j P+\Od{c^{-2}},
\label{CCE}
\n 
which can be directly derived from \Eq{POT} for $\al=0$ and $\al=j$ to corresponding order after inserting the continuity equation \Eq{CE}. Equipped with the transformation relations in Eqs. (\ref{STF}-\ref{PI4}), we are ready to solve \Eq{PF} and obtain the corresponding parameter constraints.

Substituting Eqs. (\ref{Tab}) and (\ref{Cabc}) into \Eq{PF} and taking advantage of Eqs. (\ref{STF}--\ref{PI4}), for $\al=0$, we have 
\m
&&\pa_{\bt}\[\(1+\frac{1}{c^2}A U+\frac{1}{c^4}(B U^2+\Psi_4)\)t^{0\bt}\]\notag\\
=&&\frac{1}{c}\frac{1}{4\pi}\pa_t\left\{\hf(6\gm-2A-5)\pa_jU\pa_j U\right\}+\frac{1}{c}\frac{1}{4\pi}\pa_j\left\{-(3\gm-A-2)\pa_j U\pa_t U+ 2(3\gm-A-3)\pa_kU\pa_{[k}V_{j]}\right\}
\notag\\
&&+\frac{1}{c^3}\frac{1}{4\pi}\pa_t\left\{(-\frac{23}{2}-5A-2B+5\bt+\hf\bt_2+15\gm+6A\gm-15\gm^2+6\om)U\pa_j U\pa_j U+\pa_j \[(-\Psi_1+\frac{3}{2}\Psi_3-\Psi_4+\hf \mm)\right.\right.\notag\\
&&\qquad\qquad\quad\left.\left.-\frac{1}{2}(3+A-3\gm)\Phi_1+\frac{1}{2}(A+3-3\gm)\Phi_2-(3+A-3\gm)\Phi_3-\frac{1}{2}(3+A-3\gm)\ddot{X}\]\pa_j U\right.\notag\\
&&\qquad\qquad\quad\left.-(3+A-3\gm)\(\pa_jU\pa_t V^j-\hf \pa_j V^k\pa_k V^j\)+2\Dt(\pa_k V^j\pa_k V^j)-\frac{1}{2}(3+A-3\gm-\bt_0)(\pa_tU)^2 \right\}\notag\\
&&+\frac{1}{c^3}\frac{1}{4\pi}\pa_j\left\{-\hf(3+A-3\gm)\pa_{[k}\[8((1-\tau)V_{{j}{]}}U)+2V_{1{j]}}+4(1-2\tau)V_{2{j]}}+4V_{3{j]}}+4V_{4{j]}}-8(1-\tau)\phi_{2{j]}}+2\ddot{X}_{j]}\right.\right.\notag\\
&&\qquad\qquad\quad\left.\left.+16(1-\tau)K_{1{j]}}+12(1-\frac{4}{3}\tau)K_{2{j]}}\]\pa_kU+2\pa_{[k}V_{{j}{]}}\pa_k\(-\frac{3}{2}\Psi_1+\frac{3}{2}\Psi_3-\Psi_4+\hf \mm\)\!\right.\notag\\
&&\qquad\qquad\quad\left.-\!\Dt\pa_kV^j\pa_k(-U^2+2\Phi_2-4\Phi_4)\!+\!8\Dt\pa_kV^l\pa_{[k}(\Phi_{1{{j}{]}{l}}}+P_{2{{j}{]}{l}}})-4\Dt\(V^k\pa_jU\pa_kU-\hf V^j\pa_kU\pa_kU\)\right.\notag\\
&&\qquad\qquad\quad\left.+2(-3-2B+6\bt+6\gm-15\gm^2-2A+6A\gm+6\om-4(3+A-3\gm)\tau)U\pa_kU\pa_{[k}V_{j]}\right.\notag\\
&&\qquad\qquad\quad\left.-(3+A-3\gm)\pa_k V^j\pa_t V^k+\(\frac{19}{2}+2B-4\bt-\bt_2-\frac{27}{2}\gm+15\gm^2+\frac{9}{2}A-6\gm A+2\Dt-6\om\)U\pa_t U\pa_j U\right.\notag\\
&&\qquad\qquad\quad\left.-\pa_t \(-\Psi_1+\frac{3}{2}\Psi_3-\Psi_4+\hf \mm\)\pa_j U+\pa_j\[\hf(3+A-3\gm-3\bt_1)\Phi_1+\frac{1}{2}(3\gm-4\Dt+2\bt_2-A-3)\Phi_2\right.
\right.\notag\\
&&\qquad\qquad\quad\left.\left. +(3+A-3\gm-\bt_3)\Phi_3+(4\Dt-3\bt_4)\Phi_4+\hf(3+A-3\gm-\bt_0)\ddot{X}\]\pa_tU\right\}+\frac{1}{c^3}Q_1,
\label{C1}
\n
where $\tau$ is a new free parameter introduced during transformation for the full use of the identities in \Eq{PI2} and \Eq{PI3} , and
\m
Q_1=\pa_tU\[\hf (-3\bt_1+2\gm+1)\rs v^2-(8\pi)^{-1}(-1+2\bt-\bt_2)\pa_kU\pa_k U+(1-\bt_3)\rs\Pi+(-3\bt_4+3\gm)P\].
\n
Note that apart from the terms in $Q_1$, the solution has been arranged into a combination of time derivatives and gradients of gravitational fields with the use of transformation tricks (the representative examples are given in Appendix \ref{app2}; in particular, $\tau$ is introduced in \Eq{b5}). Therefore, in order to keep the result compatible with the expression $\pa_{\bt}\[\(1+c^{-2}A U+c^{-4}(B U^2+\Psi_4)\)t^{0\bt}\]$, $Q_1$ must vanish, or equivalently, all the coefficients in $Q_1$ must vanish, namely  
\m
\bt_1=\frac{1}{3}(2\gm+1),&&\quad \bt_2=2\bt-1,\quad \bt_3=1,\quad \bt_4=\gm
\label{pr1a}.
\n 
 
For $\al=j$ in \Eq{PF}, we first solve the equation to the 1PN order, i.e., to $\Od{c^{-2}}$, 
\m
&&\pa_{\bt}\[\(1+\frac{1}{c^2}A U+\frac{1}{c^4} (B U^2+\Psi_4)\)t^{j\bt}\]\notag\\
=&&\frac{1}{4\pi}\pa_k\left\{\pa_jU\pa_kU-\dt_{jk}\(\hf \pa_l U\pa_l U\)\right\}\notag\\
&&+\frac{1}{c^2}\frac{1}{4\pi}\pa_t\left\{(8\Dt+2(A-5\gm+1))\pa_k U\pa_{[j}V_{k]}+(4\Dt-\bt_0+A-5\gm+1)(\pa_tU \pa_j U)\right\}\notag\\
&&+\frac{1}{c^2}\frac{1}{4\pi}\pa_k\left\{4\Dt\[\pa_kV^l\pa_lV^j\!+\!\pa_j V^l\pa_l V^k\!-\!\pa_j V^l \pa_k V^l\!-\!\pa_lV^j\pa_l V^k\!+\!\pa_t V^j \pa_kU\!+\!\pa_t V^k \pa_j U\!+\!\dt_{jk}\pa_l V_m\pa_{[l} V_{m]}\!-\!\dt_{jk}\pa_tV^l\pa_l U\]\right.\notag\\
&&\qquad\qquad\quad-\hf\(4\Dt\!-\bt_0\!+A\!-5\gm\!+1\)\dt_{jk}(\pa_t U\pa_t U)\!+\frac{1}{4}\pa_{(j} U\pa_{k)} \Psi_1\!-\frac{1}{2}\dt_{jk}\pa_lU\pa_l\Psi_1\notag\\
&&\qquad\qquad\quad\left.+(2\bt-A-2+5\gm-\bt_2)\[-(U\pa_kU\pa_j U)+\hf \dt_{jk}(U\pa_lU\pa_lU)\]\right. \notag\\
&&\qquad\qquad\quad+(A-5\gm+1)\[\pa_j V^l\pa_l V^k\!+\pa_t V^k \pa_j U\!+\pa_k V^j \pa_t U-\dt_{jk}\(\frac{1}{2}\pa_l V^m\pa_{m} V^{l}\!+\pa_tV^l\pa_l U\)\]\notag\\
&&\qquad\qquad\quad+(A\!-5\gm\!+1)\[\frac{1}{4}\pa_{j} U\pa_{k} (-2\Phi_2+U^2+4\Phi_4)-\frac{1}{4}\dt_{jk}\pa_lU\pa_l(-\Phi_2+U^2+2\Phi_4)\]\notag\\
&&\qquad\qquad\quad\left. +(A\!-5\gm\!+1)\[\pa_k(\Phi_1^{jl}+P_2^{jl})\pa_lU-\pa_l(\Phi_1^{jk}+P_2^{jk})\pa_lU\]\right\}+\frac{1}{c^2}Q_2^j+\Od{c^{-4}},
\label{C2}
\n
where
\m
Q_2^j=\pa_jU\[\(-\frac{1}{2}-\gm+\frac{3}{2}\bt_1\)\rs v^2+(8\pi)^{-1}(2\bt-\bt_2-1)\pa_k U\pa_kU+(-1+\bt_3)\rs \Pi+(-3\gm+3\bt_4)P\].
\label{q2j}
\n 
Similarly, requiring that all the coefficients in $Q_2$ vanish leads to the same constraints on parameters as those in \Eq{pr1a}. However, the symmetry of $t^{jk}$ has not been fully considered in the above transformation. For example, the symmetric counterpart of the term $\pa_k(\Phi_1^{jl}+P_2^{jl})\pa_lU$, namely $\pa_j(\Phi_1^{kl}+P_2^{kl})\pa_lU$, is not contained in \Eq{C2}. Here $t^{jk}$ is required to be symmetric, and hence the coefficients of such non-symmetric terms should vanish, which leads to a constraint on the constant $A$ introduced in \Eq{TET}, i.e. 
\m
A=5\gm-1.
\label{pr1b}
\n
In addition, $t^{j0}$ should also be symmetric. Comparing the time derivative part of \Eq{C2} and the gradient part of \Eq{C1} at the 1PN order, we get another two constraints,
\m
\Dt=\frac{1}{2}(\gm+1),\qquad \bt_0=1.
\label{pr1c}
\n
As a matter of fact, the solution for $\al=j$ to the 1PN order and the consequent parameter constraints have been obtained in the fundamental work for the PPN formalism by Will in \cite{Will:1971wt}, despite with some different arguments.

For $\al=j$, at the 2PN order, substituting Eqs. (\ref{Tab}) and (\ref{Cabc}) and inserting the PPN parameter constraints in Eqs. (\ref{pr1a}), (\ref{pr1b}) and (\ref{pr1c}) into \Eq{PF}, we obtain
\m
&&\pa_{\bt}\[\(1+\frac{1}{c^2}A U+\frac{1}{c^4}(B U^2+ \Psi_4)\)t^{j\bt}\]\notag\\
\supset\,
&&\cc{4}\!\left\{(\!-\!1\!-\!2B\!+\!2\bt\!-\!10\gm\!+\!15\gm^2\!+\!10\om)\rs v^j v^k U\pa_k U\!+(\!-8\Dt\!+\!\chi)\rs v^k V^j \pa_k U\!+\!\rs v^lv^k\pa_k \mm^{jl}\!+\!P\pa_k \mm^{jk}\!+\!\rs\! \mm^{jk}\pa_k U\right.\notag\\
&&\qquad+\rs v^jv^k\pa_k\(-\hf \Psi_1+\frac{5}{2}\Psi_3-\Psi_4+\hf \mm\) + \rs v^k\[\!-2\Dt v^2\!-\!(4\Dt-\chi) U\!-\!4\Dt\Pi\!-\!4\Dt P\]\pa_k V^j\!+\rs v^k\pa_k \mv^j\notag\\
&&\qquad-\hf \rs\pa_j(\mg\!+\mh\!+\mx\!+\me)\!+\!(-1\!-2B\!+2\bt\!-7\gm\!+12\gm^2\!+4\om)PU\pa_j U\!+\!(\!-1\!+2\bt)\rs \Pi U\pa_j U\notag\\
&&\qquad+\(-2-B+4\bt-5\gm+\frac{15}{2}\gm^2-2\dt+5\om\)\rs U^2\pa_j U+\rs \(-\Psi_1-\hf\Psi_2+\frac{5}{2}\Psi_3-\Psi_4+\frac{1}{2} \mm \)\pa_j U\notag\\
&&\qquad+\rs v^2\[\(\!-\frac{3}{8}\!-\frac{\gm}{2}\)v^2\!+\!\(-\frac{3}{2}\!+\bt\!-2\gm\!-2\om\)U\!+\!\(\!-\hf\!-\gm\)\Pi\!+\!(\!-1\!-\gm)P/\rs\]\pa_j U\!+\!(4\Dt-\chi)\rs v^k V^k\pa_j U\notag\\
&&\qquad+\rs v^2\pa_j\(-\frac{1}{4}\Psi_1-\hf \Psi_3\)+\rs U\pa_j\(-\hf \Psi_1-\hf \Psi_2\)+\rs \Pi \pa_j\(-\hf \Psi_1\)+P\pa_j\(-\hf \Psi_1+\Psi_3-\Psi_4\)\notag\\
&&\qquad-8\lambda \rs V^k \pa_j V^k+\rs v^k\[2\Dt v^2\!+(4\Dt-\chi) U\!+4\Dt\Pi\!+4\Dt P\]\pa_j V^k-\rs v^k \pa_j\mv^k-\hf \rs v^k v^l \pa_j\mm^{kl}\notag\\
&&\qquad+(-1-2B+2\bt-10\gm+15\gm^2+10\om)\rs U v^j \pa_t U+(-4\Dt+\chi) \rs V^j \pa_t U+\rs v^j \pa_t\(-\hf \Psi_1+\frac{5}{2}\Psi_3-\Psi_4+\hf\mm\)\notag\\
&&\qquad\left.+\rs v^k \pa_t \mm^{jk}+\rs\[-2\Dt v^2\!-(4\Dt-\chi) U\!-4\Dt\Pi\]\pa_t V^j+\rs \pa_t\mv^j\right\}.
\label{C3a}
\n 
Here $\supset$ denotes that only the terms at 2PN order are included. 
To make the transformation procedure easier, we can first determine several 2PN parameter constraints by requiring the symmetry of $t^{0j}$ at the 2PN order. For example, $\pa_j\[\(1+c^{-2}A U+c^{-4}(B U^2+\Psi_4)\)t^{0j}\]$ in \Eq{C1} contains the terms
$(4\pi c^3)^{-1}\pa_j\{-1/2(3+A-3\gm)\pa_{[k}[2V_{1{j]}}+4(1-2\tau)V_{2{j]}}+4V_{3{j]}}+4V_{4{j]}}-8(1-\tau)\phi_{2{j]}}+2\ddot{X}_{j]}+16(1-\tau)K_{1{j]}}+12(1-4/3\tau)K_{2{j]}}]\pa_kU\}$ and $(4\pi c^3)^{-1}\pa_j\{8\Dt\pa_kV^l\pa_{[k}(\Phi_{1{{j}{]}{l}}}+P_{2{{j}{]}{l}}})\}$, and then the solution of $\pa_0\[\(1+c^{-2}A U+c^{-4}(B U^2+\Psi_4)\)t^{j0}\]$ in \Eq{C3a} should correspondingly contain $(4\pi c^4)^{-1}\pa_t\{-1/2(3+A-3\gm)\pa_{[k}[2V_{1{j]}}+4(1-2\tau)V_{2{j]}}+4V_{3{j]}}+4V_{4{j]}}-8(1-\tau)\phi_{2{j]}}+2\ddot{X}_{j]}+16(1-\tau)K_{1{j]}}+12(1-4/3\tau)K_{2{j]}}]\pa_kU\}$ and  $(4\pi c^4)^{-1}\pa_t\{8\Dt\pa_kV^l\pa_{[k}(\Phi_{1{{j}{]}{l}}}+P_{2{{j}{]}{l}}})\}$, which can be obtained respectively from $\rs\pa_t \mv^{j}$ and $\rs v^k \pa_t \mm^{jk}$ in \Eq{C3a}.
Specifically, we take the transformations as follows,
\m
\rs \pa_t \mv^j&=&\frac{1}{2\pi}\pa_t(\pa_k U\pa_{[k}\mv_{j]})+\frac{1}{4\pi}\pa_k\[
 -\pa_kU\pa_t\mv^j-\pa_j U\pa_t\mv^k+\dt_{jk}(\pa_lU\pa_t \mv^l)\]\notag\\
&&+\frac{1}{4\pi}\(-\pa_t\pa_k U\pa_k \mv^j+\pa_t\pa_k U\pa_j \mv^k+\pa_j U\pa_k\pa_t \mv^k\),
\label{C3b}\\
\rs v^k \pa_t \mm^{jk}&=&\frac{1}{2\pi}\pa_t(\pa_kV^l \pa_{[k}\mm_{j]l})+\frac{1}{4\pi}\pa_k\[-\pa_kV^l\pa_t\mm^{jl}-\pa_jV^l\pa_t \mm^{kl}+\dt_{jk}(\pa_lV^m\pa_t \mm^{lm})\]\notag\\
&&+\frac{1}{4\pi}(-\pa_k\pa_t V^l\pa_k \mm^{jl}+\pa_t\pa_k V^l\pa_j \mm^{kl}+\pa_j V^l\pa_k\pa_t\mm^{kl}),
\label{C3c}
\n
and then the match between $t^{0j}$ and $t^{j0}$ provides constraints on the parameters in $\mv^j$ and $\mm^{jk}$,
\m
\chi_1\!=\!\chi_3\!=\!\chi_4\!=\!\chi_0\!=\!\hf(\gm\!+\!1),\quad \chi_2\!=\!\hf(\gm+1)(1\!-\!2\tau)&&,\quad \chi_5\!=\!\hf(\gm\!+\!1)(1\!-\!\tau),\quad \io_1\!=\!\hf(\gm\!+\!1)(1\!-\!\tau),\quad \io_2\!=\!\hf(\gm\!+\!1)(1\!-\!\frac{4}{3}\tau);\!\!\!\!\!\notag\\
\theta_1&&=\theta_2=\Dt=\hf(\gm+1).
\label{pr2a}
\n
Since $(4\pi)^{-1}\pa_j U\pa_k\pa_t \mv^k$ and $(4\pi)^{-1}\pa_j V^l\pa_k\pa_t\mm^{kl}$ on the right-hand sides of \Eq{C3b} and \Eq{C3c} satisfying the above constraints can be related to the identities of potentials in Eqs. (\ref{PI2}), (\ref{PI3}) and (\ref{PI4}), for convenience, we introduce two new expressions $\bar{\mv}^{j}$ and $\bar{\mm}^{jk}$ as follows, 
\m
\bar{\mv}^j\!&\equiv&-8(1-\tau)UV^j-2V_1^j-4(1-2\tau)V_2^j-4V_3^j-4V_4^j+8(1-\tau)\phi_2^j-2\ddot{X}^j-16(1-\tau)K_1^j-12(1-\frac{4}{3}\tau)K_2^j,\notag\\
\bar{\mm}^{jk}\!&\equiv&\!4 \Phi_1^{jk}+4P_2^{jk}.
\n
Using \Eq{pr2a} and $\bar{\mv}^j$ and $\bar{\mm}^{jk}$, after a tedious calculation, we obtain 
\m
&&\pa_{\bt}\[\(1+\frac{1}{c^2}A U+\frac{1}{c^4}(B U^2+ \Psi_4)\)t^{j\bt}\]\notag\\
\supset
\,&&\cc{4}\nor\pa_t \left\{\Dt\[ 2\pa_k((7\!-\!8\tau)U^2\!+2\Phi_1\!-\!2\Phi_2\!+4\Phi_3\!+2\ddot{X})\pa_{[j}V_{k]}+2\pa_{(t}((7-8\tau)U^2+2\Phi_1-2\Phi_2+4\Phi_3+2\ddot{X})\pa_{j)} U\right.\right.\notag\\
&&\qquad\qquad\quad\,\,\,\left.+2\pa_k U\pa_{[k}\bar{\mv}_{j]}+32(1-\tau) U\pa_k U\pa_{[k}V_{j]}+ 4(-V^k \pa_k U\pa_jU+\hf V^j \pa_k U\pa_k U-\pa_t V^k \pa_k V^j)\right.\notag\\
&&\qquad\qquad\quad\,\,\, \left.-\pa_k V^j\pa_k(2\Phi_2-U^2-4\Phi_4)-\pa_tU\pa_j(2\Phi_2-U^2-4\Phi_4)-20(1-\frac{4}{5}\tau) U \pa_t U\pa_j U+2 \pa_k V^l \pa_{[k}\bar{\mm}_{j]l}\,\] \notag\\
&&\qquad\qquad\left. +4\pa_{(t}U\pa_{j)}\(-\frac{3}{4}\mu_1\Phi_1+\hf\mu_2\Phi_2-\hf\mu_3\Phi_3-\frac{3}{2}\mu_4\Phi_4-\frac{1}{4}\mu_0 \ddot{X}\)\right\} \notag\\
&&\!+\cc{4}\!\nor\! \pa_k\!\left\{\!\Dt\!\[\!-(\pa_k V^l\pa_l\bar{\mv}^j\!+\!\pa_j V^l\pa_l\bar{\mv}^k\!+\!\pa_k\bar{\mv}^l\pa_l V^j\!+\!\pa_j\bar{\mv}^l\pa_l V^k\!-\!2\pa_{(k}V^l\pa_{j)}\bar{\mv}^l\!-\!2\pa_lV^{(k}\pa_l\bar{\mv}^{j)}\!+\!\pa_kU\pa_t\bar{\mv}^j\!+\!\pa_j U\pa_t\bar{\mv}^k)\right.\right.\notag\\
&&\qquad\qquad\quad\,\left.\left. +2\pa_{(k}((7\!-\!8\tau)U^2\!+\!2\Phi_1\!-\!2\Phi_2\!+\!4\Phi_3\!+\!2\ddot{X})\pa_t V_{j)}\!+\!\frac{1}{4}\(\pa_m\bar{\mm}^{kl}\pa_m\bar{\mm}^{jl}\!\!-\!\pa_k \bar{\mm}^{ml}\pa_m\bar{\mm}^{jl}\!\!-\!\pa_j\bar{\mm}^{ml}\pa_m\bar{\mm}^{kl}\right.\right.\right.\notag\\
&&\qquad\qquad\quad\,\,\left.\left.\!+\hf\pa_k\bar{\mm}^{ml}\pa_j\bar{\mm}^{ml}\!+\! 2\bar\mm^{jk}\pa_l U\pa_l U\!-\! 4\bar\mm^{jm}\pa_kU\pa_mU\!-\! 4\bar\mm^{km}\pa_mU\pa_jU\)\!-\!\pa_k V^l\pa_t\bar{\mm}^{jl}\!-\!\pa_jV^l\pa_t \bar{\mm}^{kl}\right.\notag\\
&&\qquad\qquad\quad\,\,\left. +\hf (\pa_{(j}\bar{\mm}_{k)l}\!-\!\pa_l\bar{\mm}_{jk})\pa_l(2\Phi_2\!\!-\!U^2\!\!-\!4\Phi_4)\!+\!2\pa_{(k}V_{j)}\pa_t(2\Phi_2\!\!-\!U^2\!+\!4\Phi_4)\!\!-\!2\pa_t V_{(k}\pa_{j)}(2\Phi_2\!\!-\!U^2\!\!-\!4\Phi_4)\right.\notag\\
&&\qquad \qquad\quad\,\,\left. +2(2\Phi_2\!\!-\!U^2\!\!-\!4\Phi_4)\pa_kU\pa_jU\!-\!\frac{1}{8}\pa_j(2\Phi_2\!\!-\!U^2\!\!-\!4\Phi_4)\pa_k(2\Phi_2\!-\!U^2\!-\!4\Phi_4)\!+\!\frac{1}{4}\pa_{(k}U^2\pa_{j)}(2\Phi_2\!\!-\!\hf U^2\!\!-\!4\Phi_4)\right.\notag\\
&&\qquad \qquad\quad\,\,\left.+\!16\!\(\!-\hf V_{(k}\pa_{j)}U\pa_t U-2(1-\tau)U\pa_{(k}V^l \pa_l V_{j)}+(1-\tau)U\pa_{k}V^l\pa_{j}V^l+(1-\tau)U\pa_lV^k\pa_lV^j\right.\right.\notag\\
&&\qquad \qquad\quad\,\,\left.\left.-2(1-\tau)U\pa_{(k}U\pa_tV_{j)}+(1-\tau)V^l\pa_{(k}V^l\pa_{j)}U+\frac{1}{4}\pa_tV^k\pa_t V^j\)\]\notag\\
&&\qquad \quad\quad +\pa_{(k}U\pa_{j)}(\mg+\mh+\me+\mx)-2\pa_{(k}\ddot{X}\pa_{j)}\(-\frac{3}{4}\mu_1\Phi_1+\hf\mu_2\Phi_2-\hf\mu_3\Phi_3-\frac{3}{2}\mu_4\Phi_4-\frac{1}{8}\mu_0 \ddot{X}\)\notag\\
&&\qquad\quad\quad-2\pa_{(k}\Phi_1\pa_{j)}\[\hf(\!-\!\frac{1}{4}\!-\!\hf\gm\!+\!\hf \om_1)\Phi_1\!+(\!-\!\hf\!+\!\bt\!+\!\om_2)\Phi_2\!+(\!-\!\hf\!-\om_3)\Phi_3\!+(\!-\!\frac{3}{2}\gm\!+\!\om_4)\Phi_4\]\!\notag\\
&&\qquad\quad\quad -2\pa_{(k}\Phi_3\pa_{j)}\[(\!-1\!+\!2\bt)\Phi_2\!-\!\hf\Phi_3\!-\!3\gm \Phi_4\]\!-\!2\pa_{(k}\Phi_2\pa_{j)}\[\hf(\!-1\!+2\bt\!-2\dt_2\!+\frac{1}{2}\Dt)\Phi_2\!+(\!-3\gm+6\dt_4\!-\Dt)\Phi_4\] \notag\\
&&\qquad\quad\quad \left.-2\pa_{(k}\Phi_4\pa_{j)}\[\hf(-B4-3\gm-2\om_4-6\Dt)\Phi_4\]\right\}\notag\\
&&\!+\cc{4}\!\nor\!\pa_j \! \left\{\!\Dt\!\[\!-(\pa_l V^m\pa_l\bar{\mv}^m-\pa_lV^m\pa_m\bar{\mv}^l-\pa_l U\pa_t \bar{\mv}^l)-\pa_l((7-8\tau)U^2+2\Phi_1-2\Phi_2+4\Phi_3+2\ddot{X})\pa_t V^l\right.\right.\notag\\
&&\qquad\qquad\quad\,-\pa_t((7-8\tau)U^2+2\Phi_1-2\Phi_2+4\Phi_3+2\ddot{X})\pa_t U+10(1-\frac{4}{5}\tau)U\pa_tU\pa_tU+\!\pa_tU\pa_t(2\Phi_2\!-\!U^2+4\Phi_4)\notag\\
&&\qquad\qquad\quad\, +\frac{1}{4}\!\(\!\hf\pa_m\bar{\mm}^{ln}\pa_n\bar{\mm}^{ml}\!\!-\!\frac{1}{4}\pa_m\bar{\mm}^{nl}\pa_m\bar{\mm}^{nl}\!\!+\! 2 \bar\mm^{ml}\pa_mU\!\pa_lU\!\!+\!4\pa_lV^m\!\pa_t \bar{\mm}^{lm}\!\!\)\!\!-\!\frac{1}{8}\pa_l U^2\pa_l(2\Phi_2\!-\!\hf U^2\!\!-\!\!4\Phi_4)\notag\\
&&\qquad\qquad\quad\, +\!16\!\(\!(1\!-\!\tau)U\!\pa_lU\!\pa_t V^l\!\!+\!\frac{1}{4}V^l\pa_lU\!\pa_t U\!\!-\!\!\hf (1\!-\!\tau)V^l\pa_m\! V^l\!\pa_m U\! \!-\!\hf (1\!-\!\tau) U\!\pa_l V^m\pa_l\!V^m\!\!+\!\hf (1\!-\!\tau) U\!\pa_l V^m\pa_m\! V^l\!\)\!\! \notag\\
&&\qquad\qquad\quad\, \left.+\pa_tV^l\pa_l(2\Phi_2\!-\!U^2\!-\!4\Phi_4)\!-\!(2\Phi_2\!-\!U^2\!-\!4\Phi_4)\pa_l U\pa_l U\!+\!\frac{3}{16}\pa_l(2\Phi_2\!-\!U^2\!-\!4\Phi_4)\pa_l(2\Phi_2\!-\!U^2\!-\!4\Phi_4)\right] \notag\\
&&\qquad\quad\quad -\hf\pa_lU\pa_l(\mg+\mh+\me+\mx)-2\pa_{t}U\pa_{t}(-\frac{3}{4}\mu_1\Phi_1+\hf\mu_2\Phi_2-\hf\mu_3\Phi_3-\frac{3}{2}\mu_4\Phi_4-\frac{1}{4}\mu_0 \ddot{X}) \notag\\
&&\qquad\quad\quad+\pa_l\ddot{X}\pa_l(-\frac{3}{4}\mu_1\Phi_1+\hf\mu_2\Phi_2-\hf\mu_3\Phi_3-\frac{3}{2}\mu_4\Phi_4-\frac{1}{8}\mu_0 \ddot{X}) \notag\\
&&\qquad\quad\quad +\pa_l\Phi_1\pa_l\[\hf(\!-\frac{1}{4}\!-\hf\gm\!+\!\hf\om_1)\Phi_1\!+(\!-\hf\!+\!\bt\!+\!\om_2)\Phi_2\!+(\!-\hf\!-\om_3)\Phi_3+(\!-\frac{3}{2}\gm\!+\!\om_4)\Phi_4\] \notag\\
&&\qquad\quad\quad\left. +\pa_l\Phi_3\pa_l\[(\!-1\!+\!2\bt)\Phi_2\!-\hf\Phi_3\!-3\gm \Phi_4\]+\pa_l\Phi_2\pa_l\[\hf(-1+2\bt-2\dt_2+\frac{1}{2}\Dt)\Phi_2+(-3\gm+6\dt_4-\Dt)\Phi_4\]\right.\notag\\
&&\qquad\quad\quad\left.+\pa_l \Phi_4\pa_l\[\hf(-B4-3\gm-2\om_4-6\Dt)\Phi_4\]\right\}+\frac{1}{c^4}Q_3^j,
\label{C4}
\n
where
\m
Q_3^j=&&\frac{1}{4\pi}\pa_j U\[6(\Dt(1-\frac{4}{3}\tau)-\xi_1)\pa_tU\pa_tU+4(\Dt-\xi_2)U\pa_t\pa_tU+8(\Dt-\xi_3)V^k\pa_k\pa_t U-8(\Dt(1-\tau)-\xi_4)\pa_kV^l\pa_l V^k\right.\notag\\
&&\qquad\qquad\, \left.+\hf \xi_5 \pa_t V^k \pa_k U+4(\Dt-\eta_1)\Phi^{kl}\pa_k\pa_lU +4(\Dt-\eta_2)P_2^{kl}\pa_k\pa_lU\]+8(\Dt(1-\tau)-\lambda)\rs V^k \pa_j V^k\notag\\
&&+(\chi-8\Dt \tau)\[-\rs v^kU(\pa_j V^k-\pa_k V^j)+\rs V^j\pa_t U+\rs U\pa_t V^j+\rs v^k V^j \pa_k U\]\notag\\
&&+\rs v^jv^k\pa_k\[-\hf \Psi_1+\frac{5}{2}\Psi_3-\Psi_4+\hf \bar{\mm}+\Dt(2\Phi_2-U^2-4\Phi_4)+\hf(-1-2B+2\bt-10\gm+15\gm^2+10\om)U^2\]\notag\\
&&+\rs v^j\pa_t\[-\hf \Psi_1+\frac{5}{2}\Psi_3-\Psi_4+\hf \bar{\mm}+\Dt(2\Phi_2-U^2-4\Phi_4)+\hf(-1-2B+2\bt-10\gm+15\gm^2+10\om)U^2\]\notag\\
&&+\pa_j U\left\{\rs\[(-1-B_1-2\gm+3\dt_1-\frac{5}{2}\om_1+2\Dt-\frac{3}{2}\kp_1)\Phi_1+(-2-B_2+4\bt-2\dt_2-5\om_2+6\Dt-\kp_2)\Phi_2\right.\right.\notag\\
&&\qquad\qquad\quad\left.\left.+(\!-\!2\!-\!B_3\!+2\dt_3\!+5\om_3\!-\!\kp_3)\Phi_3+(\!-\!B_4\!-\!6\gm\!+6\dt_4\!-\!5\om_4\!+\kp_4\!-\!8\Dt)\Phi_4+(\!-\!B_0\!-1\!+\dt_0\!+\frac{5}{2}\om_0\!-\!\hf \kp_0)\ddot{X}\]\right.\notag\\
&&\qquad\qquad\rs v^2\[(\!-\frac{3}{8}\!-\frac{1}{2}\gm+\frac{7}{8}\s_1)v^2+(\!-\frac{3}{2}+\bt\!-2\gm\!-2\om+\frac{9}{2}\s_2)U+(\!-\hf\!-\gm+\frac{3}{2}\s_4)\Pi+(\!-1\!-\gm+2\s_5)P/\rs\]\notag\\
&&\qquad\qquad+\rs U\[(-2-B+4\bt-5\gm+\frac{15}{2}\gm^2-2\dt+5\om-3\Dt+\frac{3}{2}\kp)U+(-1+2\bt-\s_6)\Pi\right.\notag\\
&&\qquad\qquad\qquad\quad\left.\left.+(-1-2B+2\bt-7\gm+12\gm^2+4\om+6\s_7-4\Dt)P/\rs\]+(4\Dt-\chi-4\s_3)\rs v^lV^l\right\}\notag\\
&&+\rs v^2\pa_j\[(-\frac{1}{4}-\hf \om_0+\frac{3}{4}\mu_1)\ddot{X}\]+\rs \Pi\pa_j\[(-\gm+\om_3)\Phi_1+(-\hf+\hf\mu_3)\ddot{X}\]\notag\\
&&+\rs U\pa_j\[(-\gm+3\dt_1-\bt-\om_2)\Phi_1+(2\dt_3-2\bt)\Phi_3+(-\hf+\dt_0-\hf\mu_2)\ddot{X}\]\notag\\
&&\!+\!P\pa_j\!\!\[\!(\!-\frac{1}{2}\!\!-\!B_1\!\!-\!\gm\!\!-\!\om_1\!\!+\!\frac{3}{2}\gm\!\!-\!\!\om_4\!)\Phi_1\!\!+\!\!(\!-\!1\!\!-\!\!B_2\!+\!\!2\bt\!\!-\!
\!2\om_2\!\!+\!4\Dt\!\!+\!3\gm\!\!-\!\!6\dt_4\!)\Phi_2\!\!+\!\!(\!-1\!\!-\!\!B_3\!+\!\!2\om_3\!\!+\!3\gm\!)\Phi_3\!\!+\!(\!-\hf\!-\!\!B_0\!\!+\!\!\om_0\!\!+\!\frac{3}{2}\mu_4\!)\ddot{X}\!\]\notag\\
\label{q3j}
\n 
Here $\supset$ also denotes that only the terms at 2PN order are included, and we guarantee the symmetry of $t^{jk}$ during transformation. Similarly, requiring that all of the coefficients in $Q_3^j$ vanish and matching the time derivative part of \Eq{C4} with the spatial derivative part of \Eq{C1} at the 2PN order, we finally obtain 
\m
&&\chi=4(1+\gm)\tau,\notag\\
&&\om_0=\gm,\qquad \om=-2(\gm+1)\tau+\bt, \qquad \om_1=\gm,\qquad \om_2=(-2\bt+2)+\gm,\qquad \om_3=\gm,\qquad \om_4=2\gm-1;\notag\\
&&\mu_0=1,\qquad \mu_1=\frac{1}{3}(2\gm+1),\qquad \mu_2=2\bt-1,\qquad \mu_3=1,\qquad \mu_4=\gm,\notag\\
&&\dt_0=\bt,\qquad \dt_1=\frac{1}{3}(2\gm-\bt+2),\qquad \dt_3=\bt,\qquad \dt_4=\gm-\bt+1,\qquad \lambda=\frac{1}{2}(\gm+1)(1-\tau),\notag\\
&& \s_1=\frac{1}{7}(4\gm+3), \qquad \s_2=\frac{1}{9}(4\gm-8(1+\gm)\tau+2\bt+3),\qquad \s_3=\hf(\gm+1)(1-2\tau),\qquad \s_4=\frac{1}{3}(2\gm+1),\notag\\
&& \s_5=\hf(\gm+1),\qquad \s_6=2\bt-1,\qquad \s_7=\hf(\gm^2-\gm-4(\gm+1)\tau+2\bt), \notag\\
&&\kp_0=2\bt-1, \quad \kp=\frac{1}{3}(\gm\!-6\bt\!+4\dt\!+4),\quad \kp_1=\frac{1}{3}(2\gm\!-2\bt\!+3), \quad \kp_2=\gm\!+2\bt\!-2\dt_2, \quad \kp_3=2\bt\!-1,\quad \kp_4=\!-\gm\!+6\bt\!-4,\notag\\
&&\xi_1=\frac{1}{2}(\gm+1)(1-\frac{4}{3}\tau),\qquad \xi_2=\xi_3=\frac{1}{2}(\gm+1),\qquad \xi_4=\frac{1}{2}(\gm+1)(1-\tau),\qquad \xi_5=0,\qquad \eta_1=\eta_2=\frac{1}{2}(\gm+1),
\label{pr2b}
\n
and
\m
&& B_0=\frac{1}{2}(5\gm-1),\quad B=\hf(15\gm^2+12\bt-12\gm-20(\gm+1)\tau-3),\quad 
B_1=-\hf(5\gm-1),\quad\notag\\ &&B_2=3(-\gm+4\bt-3),\quad B_3=5\gm-1,\quad B_4=-3(5\gm-\!1).
\label{c2}
\n
After taking all of the constraints, there remain three independent 2PN parameters which are chosen as $\dt$, $\dt_2$ and $\om$. 

To summarize, there are two independent 1PN parameters ($\gamma$ and $\beta$) and three independent 2PN parameters ($\dt$, $\dt_2$ and $\om$) in the P2PN metric \Eq{M2pn} with conservation laws. Here we make a list of all the other parameters in \Eq{M2pn} in terms of the independent parameters,
\m
&&\bt_1=\frac{1}{3}(2\gm+1),\qquad \bt_2=2\bt-1,\qquad \bt_3=1,\qquad \bt_4=\gm,\qquad \bt_0=1, \qquad \Dt=\frac{1}{2}(\gm+1),\notag\\
&& \dt_1=\frac{1}{3}(2\gm-\bt+2),\qquad \dt_3=\bt,\qquad \dt_4=\gm-\bt+1,\qquad \dt_0=\bt,\qquad \lambda=\frac{1}{4}(2\gm-\bt+\om+2),\notag\\
&& \kp=\frac{1}{3}(\gm-6\bt+4\dt+4),\qquad \kp_1=\frac{1}{3}(2\gm-2\bt+3), \qquad \kp_2=\gm+2\bt-2\dt_2, \quad \kp_3=2\bt-1,\qquad \kp_4=-\gm+6\bt-4, \notag\\
&& \kp_0=2\bt-1,\qquad\s_1=\frac{1}{7}(4\gm+3), \qquad \s_2=\frac{1}{9}(4\gm-2\bt+4\om+3),\qquad \s_3=\hf(\gm-\bt+\om+1),\qquad \s_4=\frac{1}{3}(2\gm+1),\notag\\
&& \s_5=\hf(\gm+1),\qquad \s_6=2\bt-1,\qquad \s_7=\hf(\gm^2-\gm+2\om), \notag\\
&& \mu_1=\frac{1}{3}(2\gm+1),\qquad \mu_2=2\bt-1,\qquad \mu_3=1,\qquad \mu_4=\gm,\qquad \mu_0=1, \notag\\
&&\xi_1=\frac{1}{6}(3\gm-2\bt+2\om+3),\qquad \xi_2=\frac{1}{2}(\gm+1),\qquad\xi_3=\frac{1}{2}(\gm+1),\qquad \xi_4=\frac{1}{4}(2\gm-\bt+\om+2),\qquad \xi_5=0,\notag\\
&&\eta_1=\frac{1}{2}(\gm+1),\qquad \eta_2=\frac{1}{2}(\gm+1),\notag\\
&&\chi=2(\bt-\om),\qquad
\chi_1=\hf(\gm+1),\qquad \chi_2=\hf(\gm-\bt+\om+1),\qquad
\chi_3=\hf(\gm+1),\qquad
\chi_4=\hf(\gm+1),\notag\\
&&
\chi_5=\frac{1}{4}(2\gm-\bt+\om+2),\qquad
\chi_0=\hf(\gm+1),\qquad
 \io_1=\frac{1}{4}(2\gm-\bt+\om+2),\qquad \io_2=\frac{1}{6}(3\gm-2\bt+2\om+3),\notag\\
&&\om_1=\gm,\qquad \om_2=(-2\bt+2)+\gm,\qquad \om_3=\gm,\qquad \om_4=2\gm-1,\qquad\om_0=\gm,\notag\\
&&\theta_1=\hf(\gm+1),\qquad\theta_2=\hf(\gm+1).
\label{apc}
\n
and all the constants in the total energy-momentum pseudotensor $\Theta^{\al\bt}$ \Eq{TET} read
\m
A&&=5\gm-1,\notag\\
B=\hf(15\gm^2+2\bt-12\gm+10\om-3),\quad 
B_1&&=-\hf(5\gm-1),\qquad B_2=3(-\gm+4\bt-3)\notag\\ 
B_3=5\gm-1,\quad B_4&&=-3(5\gm-\!1),\qquad B_0=\frac{1}{2}(5\gm-1).
\n

Before closing this subsection, we take some concrete examples into account. \\
(1) In GR, we have $\gm=\bt=\om=\dt=\dt_2=1$. And the solution for $\(1+\frac{1}{c^2}A U+\frac{1}{c^4} (B U^2+\Psi_4)\)t^{\al\bt}$ in Eqs.\,(\ref{C1}), (\ref{C2}) and (\ref{C4}) is exactly $(-\tilde{g})t_{LL}^{\al\bt}$ to the 2PN order given in \cite{Pati:2000vt}.\\
(2) In the scalar tensor theories \cite{Mirshekari:2013vb}, the action is given by 
\m
S=(16\pi)^{-1}\int [\phi \hat{R}-\phi^{-1}\hat{\om}(\phi)\hat{g}^{\mu\nu}\pa_\mu \phi \pa_\nu \phi]\sqrt{-\hat{g}}d^4 x+S_m(\mathfrak{m}), \hat{g}_{\mu\nu}),
\n
where $\hat{R}$ is the Ricci scalar of the spacetime metric $\hat{g}_{\mu\nu}$, $\hat{\om}(\phi)$ is an arbitrary function of the scalar field $\phi$, $S_m$ only involves the matter fields $\mathfrak{m}$ and the metric, and $\phi$ does not couple to the matter directly. We introduce the notation $\phi=\phi_0(1+c^{-2} f)$, where $\phi_0$ is asymptotic value of the scalar field far away from the system and $f$ measures the variation in $\phi$ from $\phi_0$. Then $\hat{\om}(\phi)$ can be expanded as $\hat{\om}(\phi)=\hat{\om}_0+\cc{2}\phi_0\(\frac{d\hat{\om}}{d\phi}\)_0 f+\hf \cc{4} \phi_0^2\(\frac{d^2\hat{\om}}{d\phi^2}\)_0 f^2$, where the subscript ``$0$" denotes the variable takes value at $\phi_0$. With other notations defined by 
\m
\hat{\zt}\equiv \frac{1}{4+2\hat{\om}_0},\quad \hat{\lambda}_1 \equiv \frac{(d\hat{\om}/d\phi)_0 \phi_0}{(3+2\hat{\om}_0)(4+2\hat{\om}_0)},\quad \hat{\lambda}_2 \equiv \frac{(d^2\hat{\om}/d\phi^2)_0 \phi_0^2}{(3+2\hat{\om}_0)(4+2\hat{\om}_0)^2},
\n 
we find the independent parameters given in this paper are
\m
&&\gm=1-2\hat{\zt},\qquad  \bt=1+\hat{\zt}\hat{\lambda}_1,\quad\notag\\ \om=1+\hat{\zt}(-4+4\hat{\zt}+\hat{\lambda_1}),\quad &&\dt=1-\hat{\zt}(-3\hat{\lambda}_1+\hat{\zt}\hat{\lambda}_1-4\hat{\lambda}_1^2+\hat{\lambda}_2),\quad \dt_2=1+\hat{\lambda}_1\hat{\zt}(3+2\hat{\lambda}_1).
\n 
(3) In the tensor-multiscalar theories \cite{Damour:1995kt}, Damour and Esposito-Far\`ese found there are two independent 2PN parameter labeled as $\epsilon$ and $\zeta$ via a field-theory approach, and the 2PN deviations from GR caused by two of our new parameters $\dt$ and $\dt_2$ and their related potentials $U^3$, $U\Phi_2$, $\Si(U^2)$ and $\Si(\Phi_2)$ are given by 
\m
\Dt g_{00}&=&\cc{6}\[\frac{4(\dt-1)}{3}U^3+4(\dt-1)\Si(U^2)\]+\cc{6}\[4(\dt_2-1)U\Phi_2+4(\dt_2-1)\Si(\Phi_2)\]+\Od{\frac{\gm-1}{c^6},\frac{\bt-1}{c^6},\frac{\om-1}{c^6}},\notag\\
\Dt g_{0j}&=&\Od{\frac{\gm-1}{c^5},\frac{\bt-1}{c^5},\frac{\om-1}{c^5}},\notag\\
\Dt g_{jk}&=&\Od{\frac{\gm-1}{c^4},\frac{\bt-1}{c^4},\frac{\om-1}{c^4}};
\label{CTM}
\n
then the parameters $\epsilon$ and $\zeta$ adopted in \cite{Damour:1995kt} are related to $\delta$ and $\delta_2$ by $4(\dt-1)=\epsilon$ and $2(\dt_2-1)=\zeta$.

\subsection{Generalized harmonic gauge}\label{sec3d}

After introducing these PN parameters, the harmonic gauge used in GR is no longer valid. However, Will proposed a generalized harmonic gauge matched with his PPN metric in \cite{Will:2018bme}, and we suppose such a gauge can be similarly extended to our P2PN metric in \Eq{M2pn} with all the parameter constraints \Eq{apc}.

We assume that the generalized harmonic gauge to the 2PN order takes the form, 
\m
\pa_\mu\left\{\[1+\cc{2}C\, U+\cc{4}(D\, U^2+D_1 \Phi_1+D_2 \Phi_2+D_3 \Phi_3+D_4 \Phi_4+D_0 \ddot{X})\](\sqrt{-g}g^{\mu\nu})\right\}=0,
\label{GHG}
\n
where $C$, $D$ and $D_k$ ($k=0, 1,..., 4$) are constants to be determined. With the help of identities of potentials in Eqs. (\ref{PI2}), (\ref{PI3}) and (\ref{PI4}), we find that \Eq{GHG} automatically vanish if these constants satisfy
\m
C&&=-\gm+1,\notag\\
D=\frac{1}{2}(3\gm^2-2\gm-2\bt-2\om+3),\quad D_1&&=\hf(\gm-1),\quad D_2=3-4\bt+\gm,\notag\\
D_3=-\gm+1,\quad D_4&&=3(\gm-1),\quad D_0=\hf(-\gm+1).
\n
For $\gm=\bt=\om=1$, all the above constants equal to zero and \Eq{GHG} go back to the genuine harmonic gauge.

\section{Summary and Discussion}\label{sec4}
In this paper, going along the way which Will parameterize the 1PN metric and obtain the parameter constraints by requiring some conservation laws, we extend to parameterize the 2PN metric and obtain the corresponding parameter constraints. We choose to parameterize the 2PN metric in the harmonic gauge, and the final metric meeting the constraints is proven to be in a generalized harmonic gauge. It turns out that three 2PN parameters, $\om$, $\dt$ and $\dt_2$, appear independently under the restriction of conservation laws, which enables our framework to encompass, for example, the scalar-tensor theories properly. In addition, our calculations are also in consistency with the tensor-multiscalar theories when considering the deviation from GR caused by $\dt$ and $\dt_2$.

Another important issue we need to note is about the gravitational tests. Within the PPN framework, the solar system tests put tight bounds on parameters $\gm$ and $\bt$: $|\gm-1|\leq {\cal O}(10^{-5})$ (by time delay) and  $|\bt-1|\leq {\cal O}(10^{-5})$ (by perihelion shift of Mercury) \cite{Will:2014kxa}. When it comes to the 2PN order,  Damour and Esposito-Far\`ese investigated the experimental tests for the only two independent parameters ($\epsilon$ and $\zeta$) in the tensor-multiscalar theories. Since these two parameters only appear in $g_{00}$ at $\Od{c^{-6}}$, they concluded that the light-deflection and time-delay experiments to second order cannot probe any 2PN deviation from GR \cite{Damour:1995kt}. However, the appearance of another 2PN parameter $\om$ in our present framework makes the problem more complicated, because $\om$ is involved in $g_{ij}$ at $\Od{c^{-4}}$  and possibly enter into the equation of motion of light \cite{Epstein:1980dw, Klioner:2010pu}.  \cite{Damour:1995kt} also pointed out other weak-field tests, like perihelion shift and Nordtvedt effect, are difficult to give effective access to 2PN parameters because the 2PN contribution blends in with the high-precision of 1PN parameters. On the other hand, the binary-pulsar experiments \cite{Damour:1995kt} leads to significant limits $\epsilon\leq {\cal O}(10^{-2})$ and $\zeta\leq {\cal O}(10^{-3})$ in the tensor-multiscalar theories, and other tests in the strong-field regime, like the gravitational-wave \cite{Carson:2020iik} and black-hole-shadow \cite{Psaltis:2020lvx} experiments, can also provide bounds on the 2PN parameters for a parameterized Kerr metric. It is likely that the combination of gravitational tests from different scales would eventually reveal us the nature of the gravitational theory.

Moreover, the present P2PN metric meeting the constraints is constructed under the fairly stringent conservation conditions; for a broader P2PN framework that permits the preferred-reference frames, we need other new 2PN parameters and potentials to describe the effects when the coordinates transform from the universal rest-frame to the moving frames relative to it. From an intuitive angle, there will be more than forty-nine 2PN parameters (see Eqs. (\ref{M2pn}) and (\ref{P2pn})) in a more general P2PN framework. Compared to the 1PN potentials, the 2PN potentials are more in quantity but smaller in magnitude, and the corresponding parameters' individual or joint effects remain to be investigated. We will leave this for future work.

\textit{}

{\it Acknowledgments.} 
We acknowledge the use of HPC Cluster of ITP-CAS. 
This work is supported by the National Key Research and Development Program of China Grant No.2020YFC2201502, grants from NSFC (grant No. 11975019, 11690021, 11991052, 12047503),  Strategic Priority Research Program of Chinese Academy of Sciences (Grant No. XDB23000000, XDA15020701), and Key Research Program of Frontier Sciences, CAS, Grant NO. ZDBS-LY-7009.

\appendix
\section{The metric to 2PN order in GR with regard to perfect fluid}\label{app1}
Pati and Will calculate the near-zone metric to 3.5 PN order via direct integration of the relaxed Einstein equations in \cite{Pati:2000vt}. Here we sumarize the main results for the metric to 2PN order (see \cite{Pati:2000vt} for details), 
\m
\tilde{g}_{00}&=&-(1-\hf N+\frac{3}{8}N^2-\frac{5}{16}N^3)+\hf B(1-\hf N)+\hf K^j K^j+\Od{c^{-7}},\notag\\
\tilde{g}_{0j}&=&-K^j(1-\hf N)+\Od{c^{-6}},\notag\\
\tilde{g}_{ij}&=&\dt^{ij}(1+\hf N-\frac{1}{8}N^2)+B^{ij}-\hf B\dt^{ij}+\Od{c^{-5}},\notag\\
-\tilde{g}&=&1+N-B+\Od{c^{-5}},
\label{gab}
\n
in which  
\m
N=&&\frac{4}{c^2}U_\s+\cc{4}(7U_\s^2-4\Phi_{1\s}+2\Phi_{2\s}+2\ddot{X_\s})+\cc{6}\(-16U_\s\Phi_{1\s}+8 U_\s\Phi_{2\s}+7U_\s\ddot{X_\s}+\frac{20}{3}U_\s^3-4V_\s^jV_\s^j-16\Si_\s(\Phi_{1\s})\right.\notag\\
&&\left.+\Si_\s(\ddot{X_\s})+8\Si_\s^j(V^j_\s)-2\ddot{X}_{1\s}+\ddot{X}_{2\s}+\frac{1}{6}\mathop{\ry_\s}\limits^{(4)}-4G_{1\s}-16G_{2\s}+32G_{3\s}+24G_{4\s}-16G_{5\s}-16H_{1\s}-16H_{2\s}\),\notag\\
B=&&\frac{1}{c^4}(U_\s^2+4\Phi_{1\s}-2\Phi_{2\s})+\cc{6}(U_\s\ddot{X_\s}+4V_\s^jV_\s^j-\Si_\s(\ddot{X_\s})-8\Si_\s^j(V^j_\s)+16\Si_\s^{ii}(U_\s)+2\ddot{X}_{1\s}-\ddot{X}_{2\s}-20G_{1\s}\notag\\
&&+8G_{4\s}+16G_{5\s}),\notag\\
K^j=&&\frac{4}{c^3}V_\s^j+\cc{5}(8V_{2\s}^j-8\phi_{2\s}^j+8U_\s V_\s^j+16K_{1\s}^j+12K_{2\s}^j+2\ddot{X}_{\s}^j),\notag\\
B^{ij}=&&\cc{4}\[4\Phi_{1\s}^{ij}+4P_{2\s}^{ij}-\dt^{ij}(2\Phi_{2\s}-U_\s^2)\].
\label{nbk}
\n
Here the potentials with the subscript ``$\s$" are defined by the provisional ``densities" for unspecified matter fields $\tilde{T}^{\al\bt}$, namely
\m
\s\equiv c^{-2}(\tilde{T}^{00}+\tilde{T}^{kk}),\qquad \s^{j}\equiv c^{-1}\tilde{T}^{0j}, \qquad \s^{ij}\equiv \tilde{T}^{ij},
\label{pd}
\n
through the integrals
\m
\Si_\s(f)&\eq& \int \frac{\s{'}f'}{\po} d^3x'=\rp(4\pi \s f),\qquad \qquad \quad \Si_\s^j(f)\eq \int \frac{\s^j{'} f'}{\po} d^3x'=\rp(4\pi \s^j f), \notag \\ 
\Si_\s^{ij}(f)&\eq& \int \frac{\s^{ij}{'}f'}{\po} d^3x'=\rp(4\pi \s^{ij} f), \notag\\
\rx_\s(f)&\eq&  \int \s{'} f'\po d^3x'=\rS(4\pi\s f),\qquad \quad \, \rx_\s^j(f)\eq  \int \s^j{'} f'\po d^3x'=\rS(4\pi\s^j f),\notag\\
\rx_\s^{ij}(f)&\eq&  \int \s^{ij}{'} f'\po d^3x'=\rS(4\pi\s^{ij} f),\notag\\
\ry_\s(f)&\eq&  \int\s{'}f'\spo d^3x'=\rsd(4\pi \s f);
\label{Sspc}
\n
specifically, they are given by
\m
&&U_\s\equiv \Si_\s(1),\notag\\
&&V^j_\s\equiv \Si_\s^j(1),\quad \Phi_{1\s}\equiv \Si_\s^{ii}(1),\quad \Phi_{2\s}\equiv \Si_\s(U_\s),\quad X_\s\eq\rx_\s(1),\notag\\
&&V_{2\s}^{j}\equiv \Si_\s^j(U),\!\quad\! \phi_{2\s}^j\equiv \Si_\s(V_\s^j),\! \quad\!\! K_{1\s}^j\eq\rp(U_\s^{,k}V_\s^{k,j}),\!\quad\!\! K_{2\s}^j\eq\rp(U_\s^{,j}\dot{U}_\s),\!\quad\!\! X_\s^j\equiv \rx_\s^j(1),\!\quad\!\! \Phi_{1\s}^{ij}\equiv \Si_\s^{ij}(1),\!\quad\!\! P_{2\s}^{ij}\eq P(U_\s^{,i}U_\s^{,j}),\!\!\!\!\!\!\notag\\
&&  X_{1\s}\equiv \rx_\s^{ii}(1),\quad X_{2\s}\eq \rx(U),\quad  Y_{\s}\equiv Y_\s(1),\quad G_{1\s}\eq \rp(\dot{U}_\s^2),\quad G_{2\s}\eq\rp(U_\s \ddot{U}_\s), \quad G_{3\s}\eq-\rp(\dot{U}_\s^{,k}V_\s^k),\notag\\
&& G_{4\s}\eq\rp(V_\s^{i,j}V_\s^{j,i}),\quad\, G_{5\s}\eq-\rp(\dot{V}_\s^{k}U_\s^{,k}),\quad H_{1\s}\eq\rp(\Phi_{1\s}^{ij}U_\s^{,ij}),\quad\, H_{2\s}\eq\rp(P_2^{ij}U_\s^{,ij}).
\label{pos}
\n

When the matter field is specified to be perfect fluid with the energy-momentum tensor \Eq{Tab1}, we usually convert the potentials from integrals over $\s$, $\s^j$ and $\s^{ij}$ to integrals over the conventional conserved mass density $\rs=\sqrt{-\tilde{g}}\rho u^0/c$. In terms of $\rs$, the energy-momentum tensor takes the form
\m
\tilde{T}^{\al\bt}=\rs (1+\Pi/c^2)(\sqrt{-\tilde{g}})^{-1}(u^0/c)v^\al v^\bt
+P(u^0/c)^2/c^2 v^\al v^\bt+P \tilde{g}^{\al\bt},
\label{Tab2}
\n
where $v^{\al}=(c, \bv)$ and $u^0/c=(\sqrt{-\tilde{g}_{00}-2\tilde{g}_{0j}v^j/c-\tilde{g}_{ij}v^iv^j/c^2})^{-1}$.
Combining Eqs. (\ref{pd}) and (\ref{Tab2}) as well as the metric \Eq{gab} and the functions in \Eq{nbk}, we express the provisional ``densities" in terms of $\rs$  to the required order,
\m
\s=&&\rs\[1+\cc{2}\(\frac{3}{2}v^2-U_\s+\Pi+3P/\rs\)+\cc{4}\(\frac{7}{8}v^4+\hf U_\s v^2+\frac{3}{2}\Pi v^2+2P/\rs v^2-4v^i V_\s^i+\frac{3}{2}U_\s^2-U_\s\Pi\right.\right.\notag\\
&&\quad\,\left.\left.-6U_\s P/\rs+4\Phi_{1\s}-2\Phi_{2\s}-\hf\ddot{X}_\s\)\],\notag\\
\s^{j}=&&\rs v^j\[1+\cc{2}\(\hf v^2-U_\s+\Pi+P/\rs\)\],\notag\\
\s^{ij}=&&\rs v^iv^j+\dt^{ij}P,\notag\\
\s^{ii}=&&\rs v^2+3 P+\cc{2}\[\rs v^2\(\hf v^2-U_\s+\Pi+P/\rs\)-6 U_\s P\].
\n
Substituting these formulas into the definitions in \Eq{pos} and iterating successively, we obtain the conversion relationships between the old potentials defined by $\s$, $\s^j$ and $\s^{ij}$ and the new potentials defined by $\rs$ to required order: for the Newtonian potential and 1PN potentials,
\m
U_\s=&&U+\cc{2}\(\frac{3}{2}\Phi_1-\Phi_2+\Phi_3+3\Phi_4\)+\cc{4}\(\frac{7}{8}\Si(v^4)+\hf \Si(v^2 U)-4\Si(v^i V^i)+\frac{3}{2}\Si(U^2)+\frac{5}{2}\Si(\Phi_1)-\Si(\Phi_2)\right. \notag\\
&&\left.-\Si(\Phi_3)+9\Si(\Phi_4)-\hf\Si(\ddot{X})+\frac{3}{2}\Om(v^2)+2T(v^2)-\Om(U)-6T(U)\),\notag\\
V_\s^j=&&V^j+\cc{2}\(\hf V_1^j-V_2^j+V_3^j+V_4^j\),\notag\\
\Phi_{1\s}=&&\Phi_1+3\Phi_4+\cc{2}\(\hf \Si(v^4)-\Si(v^2U)+\Om(v^2)+T(v^2)-6T(U)\),\notag\\
\Phi_{2\s}=&&\Phi_2+\cc{2}\(\frac{3}{2}\Si(\Phi_1)-\Si(\Phi_2)+\Si(\Phi_3)+3\Si(\Phi_4)+\frac{3}{2}\Si(v^2U)-\Si(U^2)+\Om(U)+3T(U)\),\notag\\
X_\s=&&X+\cc{2}\(\frac{3}{2}X_1-X_2+X_3+3X_4\);
\n
for some 2PN potentials,
\m
&&\Phi_{1\s}^{ij}=\Phi_1^{ij}+\dt^{ij}\Phi_4,\quad\,U_\s \Phi_{1\s}=U \Phi_1+3U\Phi_4,\quad\,\Si_\s(\Phi_{1\s})=\Si(\Phi_1)+3\Si(\Phi_4),\notag\\
&&X_{1\s}=X_1+3X_4,\quad\,H_{1\s}=H_1-\Si(\Phi_4),\quad\,\Si_\s^{ii}(U)=\Si(v^2U)+3T(U),
\n
and the remaining 2PN potentials in \Eq{pos} keep their forms when they are defined by $\rs$, for example, $\phi_{2\s}^j=\phi_{2}^j$, $X_{2\s}=X_2$. Substituting the conversions into \Eq{pos}
for the metric \Eq{gab}, we finally obtain the 2PN metric with regard to perfect fluid, i.e. \Eq{M2pn} with $\chi=\xi_5=0$ and the other unspecified parameters are all equal to 1.

\section{Transformation tricks}\label{app2}
Some skillful transformations are needed in solving  \Eq{PF}, and we list representative examples with regard to the equation for $\al=0$ in \Eq{C1}:\\
(1) the transformation most frequently used
\m
-4\pi \rs\pa_t\Phi_1=\na^2U \pa_t\Phi_1=-\pa_t (\pa_j\Phi_1\pa_j U)+2\pa_j(\pa_{(t} \Phi \pa_{j)} U)+4\pi \rs v^2\pa_t U;
\n
(2) different transformations for the same component
\m
&&
\begin{cases}
-4\pi \rs U\pa_t U&=\na^2 \Phi_2 \pa_tU=-\pa_t (\pa_j\Phi_2\pa_j U)+\pa_j(\pa_{t} U \pa_{j} \Phi_2)+\pa_jU\pa_j\pa_t \Phi_2,\\
-4\pi \rs U\pa_t U&=\hf \na^2 U\pa_t U^2=-\pa_t (U\pa_j U\pa_j U)+\pa_j(U\pa_{t} U \pa_{j} U)+U\pa_jU\pa_j\pa_t U,
\end{cases} \\
&&
\begin{cases}
-4\pi \rs V^j\pa_j U&=\na^2\Phi_2^{j}\pa_j U=-\pa_j(\pa_kU\pa_{[k}\phi_{2j]})+\pa_kU\pa_k\pa_j \phi_2^j,\\
-4\pi \rs V^j\pa_j U&=V^j\na^2U\pa_jU=\pa_k(V^j\pa_kU\pa_jU)-\hf\pa_j(V^j\pa_kU\pa_kU)-\hf\pa_tU \pa_kU\pa_kU-\pa_k V^j\pa_kU\pa_jU;
\end{cases}
\n
(3) synthetical transformation by use of  identities of potentials
\m
&&-4\pi \rs V^j\pa_j U-4\pi \rs v^jv^k\pa_k V^j\notag\\
=&&\[\pa_k(V^j\pa_kU\pa_jU)-\hf\pa_j(V^j\pa_kU\pa_kU)-\hf\pa_tU \pa_kU\pa_kU\]+\pa_l\pa_l P_2^{jk}\pa_k V^j+\pa_l\pa_l \Phi_1^{jk}\pa_k V^j \notag\\
=&&\[\pa_k(V^j\pa_kU\pa_jU)-\hf\pa_j(V^j\pa_kU\pa_kU)-\hf\pa_tU \pa_kU\pa_kU\]-2\pa_l\[\pa_{k}V^j\pa_{[k}(P_{2l]j}+\Phi_{1l]j})\]+\pa_lV^j\pa_l\pa_k(\Phi_1^{jk}+P_2^{jk})\notag\\
=&&\[\pa_k(V^j\pa_kU\pa_jU)-\hf\pa_j(V^j\pa_kU\pa_kU)-\hf\pa_tU \pa_kU\pa_kU\]-2\pa_l\[\pa_{k}V^j\pa_{[k}(P_{2l]j}+\Phi_{1l]j})\]\notag\\
&&-\pa_lV^j\pa_l\pa_t V^j+\frac{1}{4}\pa_l V^j\pa_l\pa_j(2\Phi_2-U^2-4\Phi_4)\notag\\
=&&\[\pa_k(V^j\pa_kU\pa_jU)-\hf\pa_j(V^j\pa_kU\pa_kU)-\hf\pa_tU \pa_kU\pa_kU\]-2\pa_l\[\pa_{k}V^j\pa_{[k}(P_{2 l]j}+\Phi_{1l]j})\]\notag\\
&&-\hf\pa_t(\pa_l V^j\pa_l V^j)+\frac{1}{4}\pa_j\[\pa_lV^j\pa_l(2\Phi_2-U^2-4\Phi_4)\]+\frac{1}{4}\pa_l\pa_tU\pa_l(2\Phi_2-U^2-4\Phi_4),
\n
\m
&&-4\pi [(2v^2\!+\!4U\!+4\Pi\!+4P/\rs)\rs v^j\pa_jU\!-\!8\rs V^j\pa_j U\!+8\rs V^j\pa_j U\!+8\rs v^j U\pa_j U]\notag\\
=&&-4\pi [(2v^2\!+\!4(1\!-\!2\tau)U\!+4\Pi\!+4P/\rs)\rs v^j\pa_jU\!-\!8(1\!-\!\tau)\rs V^j\pa_j U\!+8(1\!-\!\tau)\rs V^j\pa_j U\!+8(1\!-\!\tau)\rs v^j U\pa_j U]\notag\\
&&-64\pi \tau \rs v^j U\pa_j U \notag\\
=&&\na^2[2V_1^j\!+4(1\!-\!2\tau)V_2^j\!+4V_3^j\!+4V_4^j\!-\!8(1\!-\!\tau)\phi_2^j\!+8(1\!-\!\tau)(UV)^j\!+16(1\!-\!\tau)K_1^j\!+12(1\!-\!\frac{4}{3}\tau)K_2^j\!+2\ddot{X}^j]\pa_jU\!\notag\\
&&+\!12(1\!-\!\frac{4}{3}\tau)\pa_kU\pa_kU\pa_tU\!-\!4\pa_t\pa_tV^j\pa_jU-16\tau \[2\pa_j(U\pa_kU \pa_{[k}V_{j]})+U\pa_kU\pa_k\pa_t U\]\notag\\
=&&-2\pa_j\{\pa_{[k}[2V_{1{j]}}\!+\!4(1\!-\!2\tau)V_{2{j]}}\!+\!4V_{3{j]}}\!+\!4V_{4{j]}}\!-\!8(1\!-\!\tau)\phi_{2{j]}}\!+\!8(1\!-\!\tau)(V_{{j}{]}}U)\!+\!16(1\!-\!\tau)K_{1{j]}}\!+\!12(1\!-\!\frac{4}{3}\tau)K_{2{j]}}\!+\!2\ddot{X}_{j]}]\pa_kU\}\notag\\
&&+\pa_k\pa_j[2V_1^j\!+4(1\!-\!2\tau)V_2^j\!+4V_3^j\!+4V_4^j\!-\!8(1\!-\!\tau)\phi_2^j\!+8(1\!-\!\tau)(UV^j)\!+16(1\!-\!\tau)K_1^j\!+12(1\!-\!\frac{4}{3}\tau)K_2^j\!+2\ddot{X}^j]\pa_kU\notag\\
&&+12(1-\frac{4}{3}\tau)\pa_kU\pa_kU\pa_tU-4\pa_t\pa_tV^j\pa_jU-16\tau \[2\pa_j(U\pa_kU \pa_{[k}V_{j]})+U\pa_kU\pa_k\pa_t U\]\notag\\
=&&-2\pa_j\{\pa_{[k}[2V_{1{j]}}\!+\!4(1\!-\!2\tau)V_{2{j]}}\!+\!4V_{3{j]}}\!+\!4V_{4{j]}}\!-\!8(1\!-\!\tau)\phi_{2{j]}}\!+\!8(1\!-\!\tau)(V_{{j}{]}}U)\!+\!16(1\!-\!\tau)K_{1{j]}}\!+\!12(1\!-\!\frac{4}{3}\tau)K_{2{j]}}\!+\!2\ddot{X}_{j]}]\pa_kU\}\notag\\
&&-\pa_k\pa_t((7-8\tau)U^2\!+\!2\Phi_1\!-\!2\Phi_2\!+\!4\Phi_3\!+\!2\ddot{X})\pa_kU\!+\!12(1-\frac{4}{3}\tau)\pa_kU\pa_kU\pa_tU\!-\!4\pa_t(\pa_tV^j\pa_jU)\!+\!2\pa_t(\pa_jV^k\pa_kV^j)\!\notag\\
&&-\!4\pa_j(\pa_k V^j\pa_t V^k)-32\tau \pa_j(U\pa_kU \pa_{[k}V_{j]})-16\tau U\pa_kU\pa_k\pa_t U\notag\\
=&&-2\pa_j\{\pa_{[k}[2V_{1{j]}}\!+\!4(1\!-\!2\tau)V_{2{j]}}\!+\!4V_{3{j]}}\!+\!4V_{4{j]}}\!-\!8(1\!-\!\tau)\phi_{2{j]}}\!+\!8(1\!-\!\tau)(V_{{j}{]}}U)\!+\!16(1\!-\!\tau)K_{1{j]}}\!+\!12(1\!-\!\frac{4}{3}\tau)K_{2{j]}}\!+\!2\ddot{X}_{j]}]\pa_kU\}\notag\\
&&\!-\!4\pa_t(\pa_tV^j\pa_jU)\!+\!2\pa_t(\pa_jV^k\pa_kV^j)-\!4\pa_j(\pa_k V^j\pa_t V^k)-32\tau \pa_j(U\pa_kU \pa_{[k}V_{j]})\notag\\
&&-\pa_k\pa_t(7U^2\!+\!2\Phi_1\!-\!2\Phi_2\!+\!4\Phi_3\!+\!2\ddot{X})\pa_kU\!
\label{b5}
\n

\bibliography{./ref}
\end{document}